\documentclass[fleqn,usenatbib]{mnras}

\usepackage{newtxtext,newtxmath}

\usepackage[T1]{fontenc}

\DeclareRobustCommand{\VAN}[3]{#2}
\let\VANthebibliography\thebibliography
\def\thebibliography{\DeclareRobustCommand{\VAN}[3]{##3}\VANthebibliography}

\usepackage{graphicx}	
\usepackage{amsmath}	
\usepackage{subcaption}
\usepackage[english]{babel}
\usepackage{multirow}
\usepackage{multicol}

\title[Chromatic optical polarization of BL Lac]{Chromatic optical polarization of BL Lac: while faint and bright}

\author[E. Shablovinskaya et al.]{
Elena Shablovinskaya,\thanks{E-mail: shabli@sao.ru, e.shablie@yandex.com}
Eugene Malygin,
Dmitry Oparin
\\
Special astrophysical observatory of Russian Academy of Sciences, Nizhnij Arkhyz, Karachai-Cherkessian Republic,
369167, Russia
}

\date{Accepted XXX. Received YYY; in original form ZZZ}

\pubyear{2022}

\begin{document}
\label{firstpage}
\pagerange{\pageref{firstpage}--\pageref{lastpage}}
\maketitle

\begin{abstract}
Due to the first results on astrophysical X-ray polarization provided by \textit{IXPE} observatory, the interest in wavelength-dependent synchrotron polarization of BL Lac type objects increases. This paper presents the results of multi-band optical observations of the well-known blazar named BL Lac ($z=0.069$) in polarized light. It was shown that the object's emission, regardless of its phase of activity, is characterized by the intraday variability of brightness and polarization with changes occurring on a time-scale of up to 1.5 hours without any stable oscillation period. Polarimetric observations in the different optical bands show that the degree and angle of polarization of the blazar depend on the wavelength, and the maximum chromatism, as well as the maximum observed polarization degree, was detected during the minimum brightness state; during the flare state, the polarization chromatism changed along with the flux gradient on the time-scale of an hour. Qualitatively, such behaviour can be described by the shock-in-jet model, yet the chromatism amplitude and its rapid changes differ significantly from the model predictions and challenge the numerical calculations.
\end{abstract}




\section{Introduction}

The different orientation of active galactic nuclei (AGN) relative to the observer allows for studying various physical manifestations of the matter accretion on to a supermassive black hole (SMBH) \citep{UnMod,Ant93}. BL Lac type objects, or blazars, are a special type of AGN, where the relativistic jet is oriented so that one can reveal the processes occurring inside the jet -- from the  regions very close to the SMBH in the short-wave spectral range to parsecs and kiloparsecs scales in radio waves.

It has long been clear that determining the polarimetric properties of blazar radiation, in particular the variability of polarization and its dependence on wavelength, can help in studying the physical processes in the emitting plasma and its dynamics, as well as the characteristics and structure of the magnetic field in a jet \citep[see e.g.][and references therein]{zhang19}. A breakthrough step there is associated with the launch of the \textit{IXPE} satellite, which has already allowed us to move much further in understanding the structure of the jet. The observed differences between optical and X-ray polarization have become an important argument in favour of the shock model in the blazar jet \citep{liodakis22}.

\begin{table}
\begin{tabular}{l|l|ccccc}
\hline
 & Date       & JD       & Device & Filter(s) & Duration & $\Delta t$ \\ \hline
1 & 22/06/2020 & 022.981 & S      & $V$ + $I$ & 95       & 8          \\
2 & 30/06/2020 & 030.946 & S      & $V$       & 140      & 1.5        \\
3 & 24/07/2020 & 054.920 & S      & $V$       & 313      & 1          \\
4 & 23/08/2020 & 084.941 & S      & $V$ + $I$ & 321      & 2          \\
5 & 24/08/2020 & 085.839 & S      & $V$ + $I$ & 109      & 2          \\
 &            &           &        & $I$       & 172      & 1          \\
6 & 24/10/2020 & 146.802 & M      & CAMEL & 260      & 2          \\
7 & 25/10/2020 & 147.785 & M      & CAMEL & 265      & 2          \\
8 & 28/06/2022 & 758.937 & M      & CAMEL & 170      & 3.5        \\
9 & 29/06/2022 & 759.908 & M      & CAMEL & 236      & 5          \\
10 & 30/06/2022 & 760.918 & M      & CAMEL & 247      & 5          \\
11 & 30/08/2022 & 821.904 & M      & SED550  & 409      & 7          \\ 
  &   &   &        & + SED650  &       &           \\ \hline
\end{tabular}
\caption{Log of observations. The date is given in dd/mm/yyyy format, JD = JD - 2459000.5. "S"\ is for the StoP device, and "M"\ is for the MAGIC focal reducer. Duration and cadence $\Delta$t are given in minutes. }
\label{log_obs}
\end{table}

The processes observed in the optical range, in particular, rapid intraday variability (IDV) of brightness and polarization in blazars, are described qualitatively by the model of shock-waves in jet \citep{marscher08,marscher16}, however, an accurate quantitative description has not yet been obtained. In addition, the possible influence of other processes -- both the contribution of external AGN regions outside the jet and additional processes inside it -- remains uncertain. In the 1980s-90s there was a special interest in the observation of polarization chromatism in the optical and near-IR ranges \citep[][and etc.]{bjo85,ballard90}. There a wide range of physical processes and models was proposed, which allows us to explain in general terms the observed dependence of synchrotron polarization on wavelength. In recent years, new models have also continued to appear \citep[see e.g. results in][]{marscher21}. Concerning the ongoing work of the \textit{IXPE} observatory, multi-wave observations of polarization within the optical range seem an important addition and test for models aimed at describing the polarization features in various spectral ranges. 

To determine the characteristics of the optical frequency-dependent polarization degree and angle, in 2020-2022 we conducted observations of the well-known object BL Lac ($z=0.069$, RA = 22 02 43.3, Dec = +42 16 40, J2000), the ancestor of the blazar objects class.
Since the object showed historical flares in 2020 and passed through a deep minimum phase in 2022, our interest was to trace how the intraday variability of brightness and polarization depends on the phase of the source activity, to assess whether the polarized radiation of the object will be characterized by chromatism, and also to track the change in these characteristics between epochs. 

This paper is arranged as follows. Section \ref{obs} describes the observations of BL Lac in 2020-2022 on the 1-m telescope of the SAO RAS; the results of these observations are given in Section \ref{res}. Section \ref{chro} gives a detailed discussion on the possible physical mechanisms of chromatism of BL Lac polarization. The conclusion is given in Section \ref{con}.

\begin{figure}
    \centering
    \includegraphics[width=0.44\textwidth]{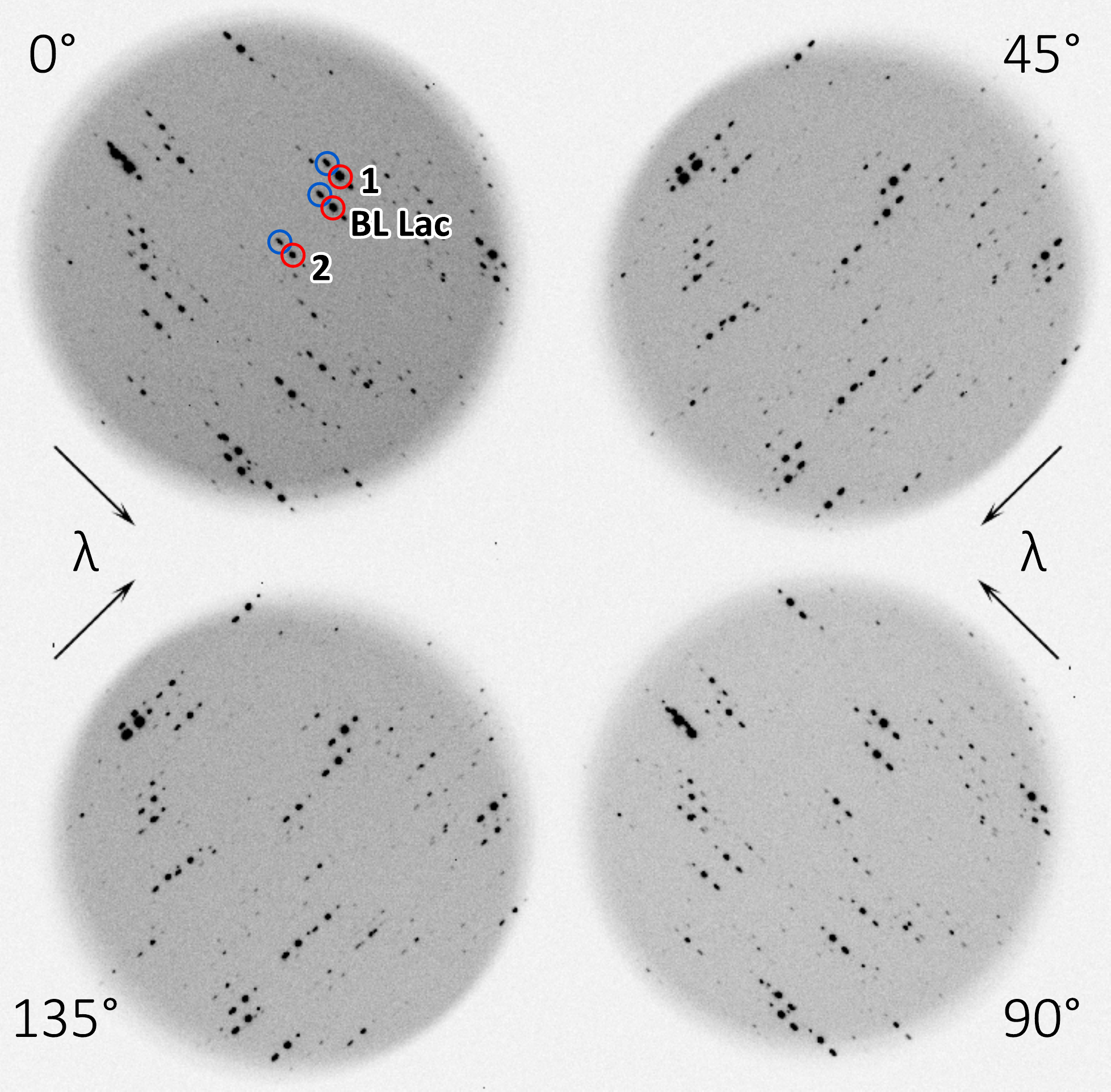}
    \caption{Image of the BL Lac field obtained at Zeiss-1000 telescope using the MAGIC device with a quadrupole Wollaston prism and a CAMEL filter. The directions of the electric vector oscillation and the direction of the prism dispersion are indicated for each of the four images. The image in the 0$^\circ$ direction shows an object and two stars -- local standards. Blue and red circles mark the "blue"{} and "red"{} images of the sources.}
    \label{camel_ima}
\end{figure}

\begin{figure}
    \centering
    \includegraphics[width=0.27\textwidth,angle=90]{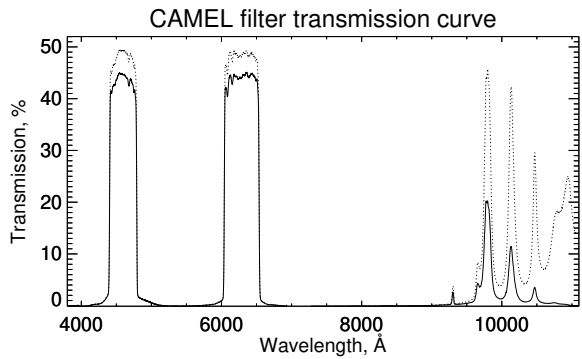}
    \caption{CAMEL filter transmission curve. The dashed line is the measured transmission curve, the solid line is the transmission curve, convolved with the CCD sensitivity curve. For observations, two peaks with CL=4600\AA{} and CL=6300\AA{} are used, separated by $10''$ in the receiver plane due to the dispersion of the quadrupole Wollaston prism. Artefacts of filter transmission in the near-IR range form a parasitic image of sources on the frame, which does not interfere with "blue"{} and "red"{} images.}
    \label{camel_trans}
\end{figure}

\section{Observations}\label{obs}

Polarimetric observations of BL Lac were carried out in summer-autumn 2020 and summer 2022 at the 1-m Zeiss-1000 telescope of SAO RAS \citep{Z1000}. Two devices were used for observations: a photometer-polarimeter StoP \citep{stop} and a focal reducer \textsc{MAGIC} \citep{magic_an}. The general log of observations is given in Table \ref{log_obs}. Data reduction included standard steps of bias and flat-fielding subtraction.

In the polarimetric mode, the StoP device uses a double Wollaston prism of the wedged design \citep{oliva}. The image of the input mask with the size $\sim$1$'\times$6$'$ was registered on the receiver in four directions of the electric vector oscillation 0$^\circ$, 90$^\circ$, 45$^\circ$ and 135$^\circ$ separated so that they do not overlap each other. In the case of MAGIC, a Wollaston prism of the original design, the so-called quadrupole prism, proposed for the first time by \citet{geyer}, is used. This prism also allows one to register four images of the input mask, which in this case is a square of  a 6.5$'\times$6.5$'$ size.

Double Wollaston prisms allow one to conduct observations in the so-called one-shot polarimetry mode unaltered by rapid atmosphere flickering. Thus, the Stokes parameters $I$, $Q$, $U$ are measured for each frame, calculated using the formulae:
\begin{gather*} 
I = I_0 + I_{90}D_{\rm Q} + I_{45} + I_{135}D_{\rm U}, \\ 
Q = \frac{I_0 - I_{90} D_{\rm Q} }{I_0 + I_{90} D_{\rm Q}} , \\
U = \frac{I_{45} - I_{135} D_{\rm U}}{I_{45} + I_{135} D_{\rm U}} ,
\end{gather*}
where $D_{\rm Q}$ and $D_{\rm U}$ are coefficients of polarization channel transmission. Here and further in the text, $Q$ and $U$ are the Stokes parameters normalized by intensity. $D_{\rm Q}$ and $D_{\rm U}$ are calculated for each frame using the local standard in the field, which allows one not only to take into account instrumental effects but also to correct atmospheric variations during observations. For more details, see \citet{AfAm}. The polarization degree $P$ and the angle of the polarization plane $\varphi$ then are equal to:
\begin{gather*}
{P} = \sqrt{Q^2 + U^2},\\
{\varphi} = \frac{1}{2} \arctan Q/U.
\end{gather*}
Note here that to transform the Stokes parameters to the celestial plane, one should multiply the Stokes vector by the rotation matrix with a $-2\cdot$PA angle, where PA is the positional angle of the device.

In the case of observations of BL Lac, we used the local standard \#1 (see Fig. \ref{camel_ima}) at a distance of $\sim$5$"$ from the object as a reference star in all epochs. The such closeness of the reference star to the source makes it possible to minimize the possible residual effects of instrumental polarization over the field, as well as to take into account the effects of ISM polarization. However, since we are not taking the average polarization of all the stars of the field, but only one star, the ISM polarization bias can be introduced into the measured value, which may depend on the colour, but be constant over time.

\begin{table*}
\begin{tabular}{ccccccccc}
    & Date       & \ \ \ \ Band                                            & \ \ \ \ \ \textless{}Mag\textgreater{}                                                & \ \ \ \ \ $F_{\rm var}$, \%                                   & \ \ \ \ \  \textless{}$P$\textgreater{}, \%                                        & \ \ \ \ \  \textless{}$\varphi$\textgreater{}, deg                                   & \textless{}$\Delta$Mag\textgreater{} &                         \\ \hline
(1) & (2)        & \multicolumn{1}{c}{ \ \ \ \ \ (3)}                           & \multicolumn{1}{c}{ \ \ \ \ \ (4)}                                                     & \multicolumn{1}{c}{ \ \ \ \ \ (5)}                             & \multicolumn{1}{c}{ \ \ \ \ \ (6)}                                                 & \multicolumn{1}{c}{ \ \ \ \ \ (7)}                                                   & \multicolumn{1}{c}{(8)}              & \multicolumn{1}{c}{(9)} \\ \hline
1   & 22/06/2020 & \begin{tabular}[c]{@{}c@{}}$V$\\ $I$\end{tabular} & \begin{tabular}[c]{@{}c@{}}14.42 $\pm$ 0.03\\ 13.02 $\pm$ 0.02\end{tabular} & \begin{tabular}[c]{@{}c@{}}9.4\\ 8.3\end{tabular}   & \begin{tabular}[c]{@{}c@{}}7.7 $\pm$ 0.5\\ 6.7 $\pm$ 0.2\end{tabular}   & \begin{tabular}[c]{@{}c@{}}48.6 $\pm$ 3.3\\ 51.5 $\pm$ 1.9\end{tabular}   & 1.4                                  & I                       \\  
2   & 30/06/2020 & \ \ \ \ \ $V$                                               & \ \ \ \ \ 13.95 $\pm$ 0.01                                                            & \ \ \ \ \ 3.1                                                 & \ \ \ \ \ 7.4 $\pm$ 0.5                                                           & \ \ \ \ \ 160.3 $\pm$ 2.0                                                           &                                      &                         \\
3   & 24/07/2020 &\ \ \ \ \ $V$                                               & \ \ \ \ \ 14.23 $\pm$ 0.04                                                            & \ \ \ \ \ 10.0                                                & \ \ \ \ \ 10.8 $\pm$ 0.9                                                          & \ \ \ \ \ 75.8 $\pm$ 2.0                                                            &                                      &                         \\ \hline
4   & 23/08/2020 & \begin{tabular}[c]{@{}c@{}}$V$\\ $I$\end{tabular} & \begin{tabular}[c]{@{}c@{}}12.69 $\pm$ 0.03\\ 11.39 $\pm$ 0.02\end{tabular} & \begin{tabular}[c]{@{}c@{}}13.8\\ 13.0\end{tabular} & \begin{tabular}[c]{@{}c@{}}2.9 $\pm$ 1.2\\ 3.3 $\pm$ 0.9\end{tabular}   & \begin{tabular}[c]{@{}c@{}}50.6 $\pm$ 12.7\\ 48.4 $\pm$ 10.1\end{tabular} & 1.3                                  & \multirow{3}{*}{II}     \\
5   & 24/08/2020 & \begin{tabular}[c]{@{}c@{}}$V$\\ $I$\end{tabular} & \begin{tabular}[c]{@{}c@{}}12.74 $\pm$ 0.02\\ 11.45 $\pm$ 0.02\end{tabular} & \begin{tabular}[c]{@{}c@{}}9.5\\ 8.4\end{tabular}   & \begin{tabular}[c]{@{}c@{}}8.9 $\pm$ 0.6\\ 7.3 $\pm$ 0.3\end{tabular}   & \begin{tabular}[c]{@{}c@{}}120.6 $\pm$ 2.8\\ 120.9 $\pm$ 2.2\end{tabular} & 1.3                                  &                         \\
    &            & \ \ \ \ \ $I$                                               & \ \ \ \ \ 11.51 $\pm$ 0.03                                                            & \ \ \ \ \ 12.0                                                & \ \ \ \ \ 5.3 $\pm$ 1.5                                                           & \ \ \ \ \ 124.0 $\pm$ 2.4                                                           &                                      &                         \\ \hline
6   & 24/10/2020 & \begin{tabular}[c]{@{}c@{}}$B$\\ $R$\end{tabular} & \begin{tabular}[c]{@{}c@{}}14.97 $\pm$ 0.02\\ 12.96 $\pm$ 0.01\end{tabular} & \begin{tabular}[c]{@{}c@{}}9.8\\ 7.1\end{tabular}   & \begin{tabular}[c]{@{}c@{}}11.0 $\pm$ 1.2\\ 11.3 $\pm$ 0.8\end{tabular} & \begin{tabular}[c]{@{}c@{}}175.0 $\pm$ 2.6\\ 172.1 $\pm$ 1.7\end{tabular} & 2.0                                  & \multirow{2}{*}{III}    \\
7   & 25/10/2020 & \begin{tabular}[c]{@{}c@{}}$B$\\ $R$\end{tabular} & \begin{tabular}[c]{@{}c@{}}14.71 $\pm$ 0.04\\ 12.73 $\pm$ 0.03\end{tabular} & \begin{tabular}[c]{@{}c@{}}15.2\\ 13.2\end{tabular} & \begin{tabular}[c]{@{}c@{}}7.2 $\pm$ 0.9\\ 8.1 $\pm$ 0.9\end{tabular}   & \begin{tabular}[c]{@{}c@{}}170.2 $\pm$ 3.6\\ 166.5 $\pm$ 3.1\end{tabular} & 2.0                                  &                         \\ \hline
8   & 28/06/2022 & \begin{tabular}[c]{@{}c@{}}$B$\\ $R$\end{tabular} & \begin{tabular}[c]{@{}c@{}}16.31 $\pm$ 0.01\\ 14.20 $\pm$ 0.01\end{tabular} & \begin{tabular}[c]{@{}c@{}}3.5\\ 3.5\end{tabular}   & \begin{tabular}[c]{@{}c@{}}26.1 $\pm$ 1.3\\ 23.5 $\pm$ 0.5\end{tabular} & \begin{tabular}[c]{@{}c@{}}16.3 $\pm$ 1.1\\ 11.3 $\pm$ 0.5\end{tabular}   & 2.1                                  & \multirow{3}{*}{IV}     \\
9   & 29/06/2022 & \begin{tabular}[c]{@{}c@{}}$B$\\ $R$\end{tabular} & \begin{tabular}[c]{@{}c@{}}16.25 $\pm$ 0.02\\ 14.15 $\pm$ 0.01\end{tabular} & \begin{tabular}[c]{@{}c@{}}4.9\\ 6.5\end{tabular}   & \begin{tabular}[c]{@{}c@{}}26.4 $\pm$ 1.3\\ 24.9 $\pm$ 0.5\end{tabular} & \begin{tabular}[c]{@{}c@{}}18.9 $\pm$ 1.1\\ 12.9 $\pm$ 0.6\end{tabular}   & 2.1                                  &                         \\
10  & 30/06/2022 & \begin{tabular}[c]{@{}c@{}}$B$\\ $R$\end{tabular} & \begin{tabular}[c]{@{}c@{}}16.35 $\pm$ 0.02\\ 14.26 $\pm$ 0.01\end{tabular} & \begin{tabular}[c]{@{}c@{}}5.1\\ 4.1\end{tabular}   & \begin{tabular}[c]{@{}c@{}}25.5 $\pm$ 1.4\\ 22.5 $\pm$ 0.8\end{tabular} & \begin{tabular}[c]{@{}c@{}}17.8 $\pm$ 1.7\\ 11.8 $\pm$ 0.8\end{tabular}   & 2.1                                  &                         \\ \hline
11  & 30/08/2022 & \begin{tabular}[c]{@{}c@{}}$V$\\ $R$\end{tabular} & \begin{tabular}[c]{@{}c@{}}14.22 $\pm$ 0.02\\ 13.52 $\pm$ 0.02\end{tabular} & \begin{tabular}[c]{@{}c@{}}6.3\\ 6.8\end{tabular}   & \begin{tabular}[c]{@{}c@{}}12.0 $\pm$ 0.8\\ 11.6 $\pm$ 0.7\end{tabular} & \begin{tabular}[c]{@{}c@{}}1.8 $\pm$ 4.8\\ 1.1 $\pm$ 4.2\end{tabular}     & 0.7                                  & V                       \\ \hline
\end{tabular}
\caption{Table of average values of brightness, amplitude of variability and polarization of BL Lac depending on the epoch and photometric band: (1) epoch number; (2) date of observations in the format dd/mm/yyyy; (3) photometric band; (4) average magnitude; (5) amplitude of variability $F_{\rm var}$ in per cent; (6) average degree of polarization <$P$> in per cent; (7) average angle of polarization <$\varphi$> in degrees; (8) average difference of magnitudes in two bands, characterizing the colour of the object; (9) the state of the object activity (see Sec. \ref{res}).}
\label{mean_val}
\end{table*}

In order for the effect of polarization chromatism to be detectable, we selected observational bands that were as far apart in terms of wavelengths from each other as possible. For observations on StoP, we chose the Johnson system broadband filters $V$ and $I$. The observations in the same broad bands with MAGIC quadrupole Wollaston prism are impossible since the prism has a significant dispersion, much greater than the images formed by seeing \citep[see][for more detailed characteristics of the prism]{magic_new}. Therefore, for all the nights of observations except the last one with the MAGIC device, we used a so-called CAMEL filter produced by Chroma Technology\footnote{\url{https://www.chroma.com /}} with two bandpasses. The  filter transmission curve is given in Fig. \ref{camel_trans}. The first band (CWL=4600\AA, FWHM=330\AA) is oriented close to the maximum Johnson $B$-band and the second band (CWL=6300\AA, FWHM=460\AA) is close to Johnson $R$-band, so further in the text we will use the notation $B$ and $R$ for this observational mode\footnote{Note also that CAMEL is medium-band (less than 500\AA{}) filter with maximum transmission less than 50\%. Thus, the filter is suitable for observations of only bright objects.}. Also in Fig. \ref{camel_trans} it can be seen that in the wavelength range $\sim$1 $\mu$m, filter transmission artifacts appear. Due to the use of a thick CCD Andor iKon-L 936 BEX2-DD characterized by high sensitivity in the red and near-IR ranges an artefact IR image of the sources is formed but does not affect the "blue"{} and "red"{} images, so we neglect it. The dispersion of the Wollaston prism separates the images in $B$ and $R$ bands at a distance of $10''$. The distortion of star-shaped images due to the prism dispersion does not exceed $3''$, which does not complicate photometry. 
In the last epoch of observations (08/30/2022), we used MAGIC with the usual wedged Wollaston prism from the StoP polarimeter device with a narrow mask of $31"\times$9$'$ size. The observations were carried out with the SED550 and SED650 medium-band filters from the SCORPIO-2 set\footnote{\url{https://www.sao.ru/hq/lsfvo/devices/scorpio-2/filters_eng.html }}, corresponding to $V$ and $R$ bands.

\section{Results}\label{res}

\begin{figure*}
     \centering
     \begin{subfigure}[b]{0.325\textwidth}
         \centering
         \includegraphics[width=\textwidth]{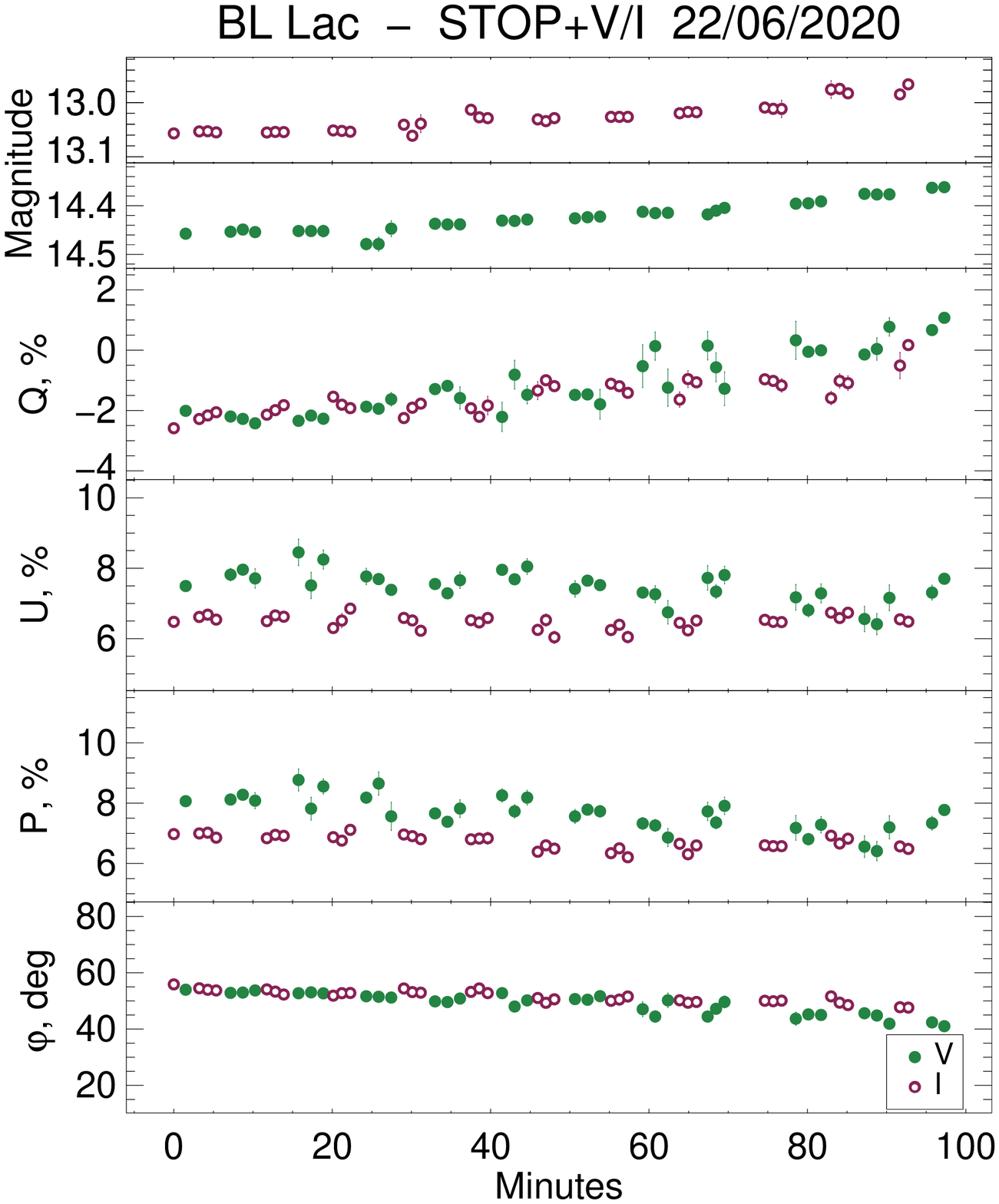}
     \end{subfigure}
     \hfill
     \begin{subfigure}[b]{0.325\textwidth}
         \centering
         \includegraphics[width=\textwidth]{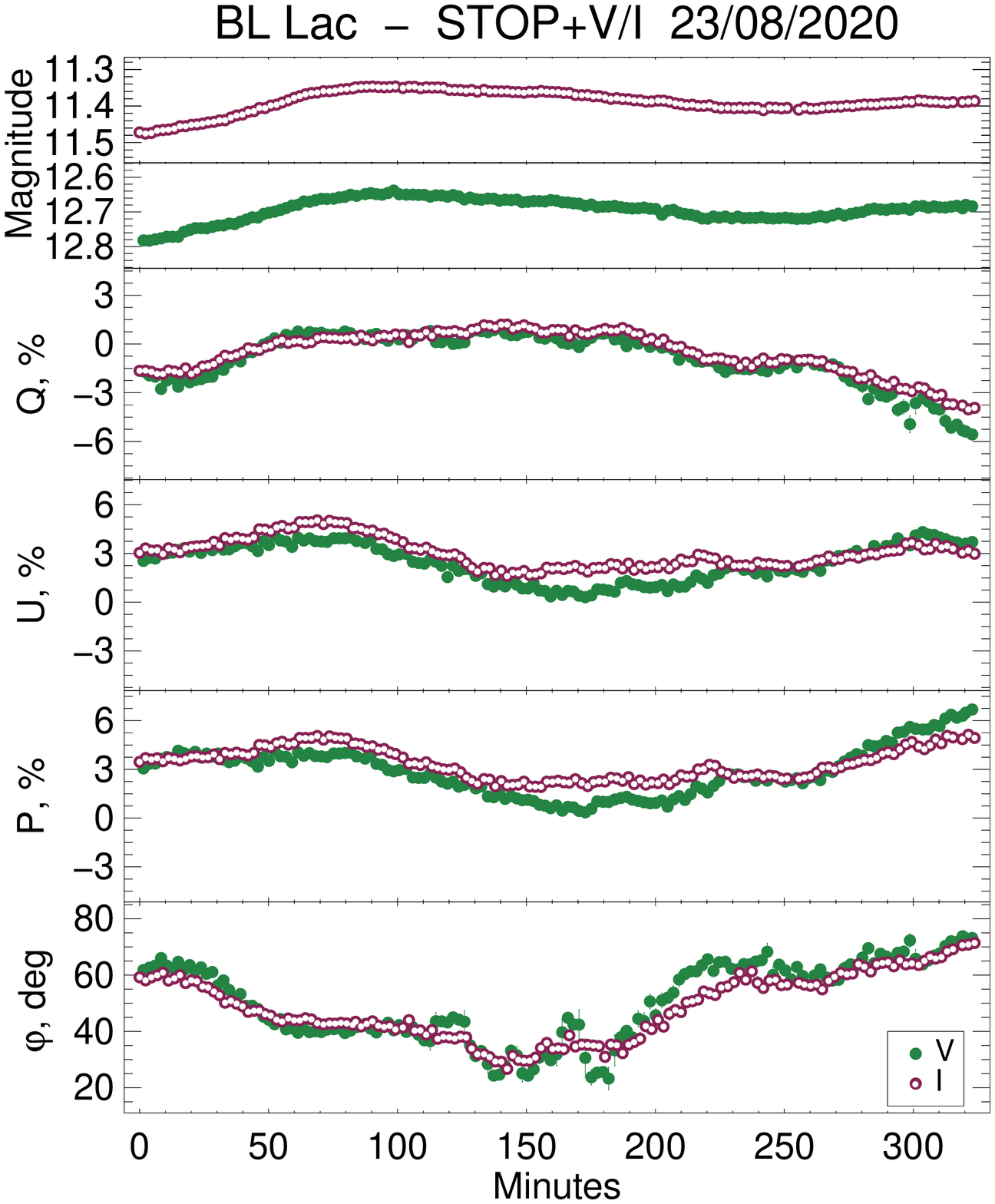}
     \end{subfigure}
     \hfill
     \begin{subfigure}[b]{0.325\textwidth}
         \centering
         \includegraphics[width=\textwidth]{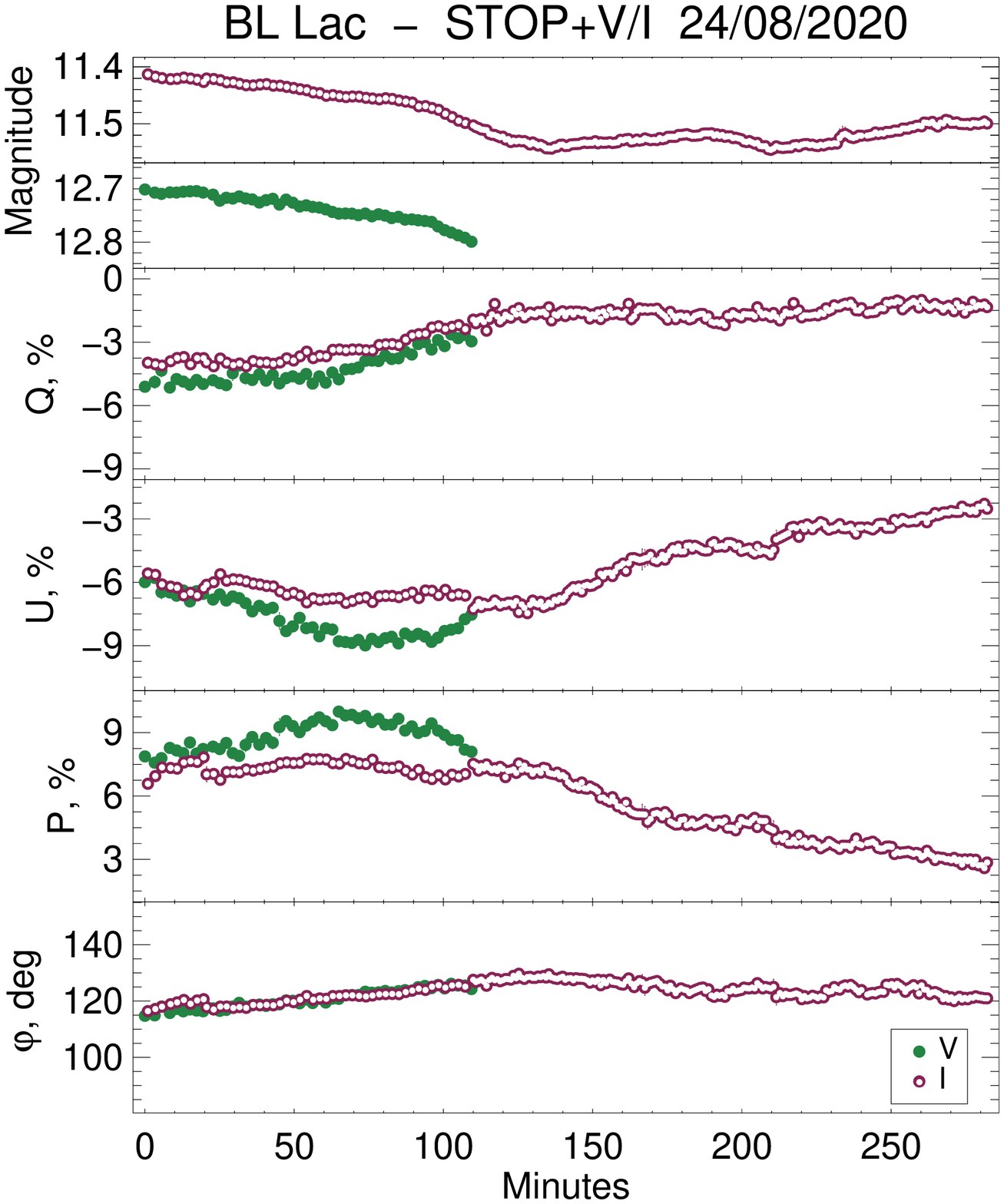}
     \end{subfigure}
    \hfill
    \begin{subfigure}[b]{0.325\textwidth}
         \centering
         \includegraphics[width=\textwidth]{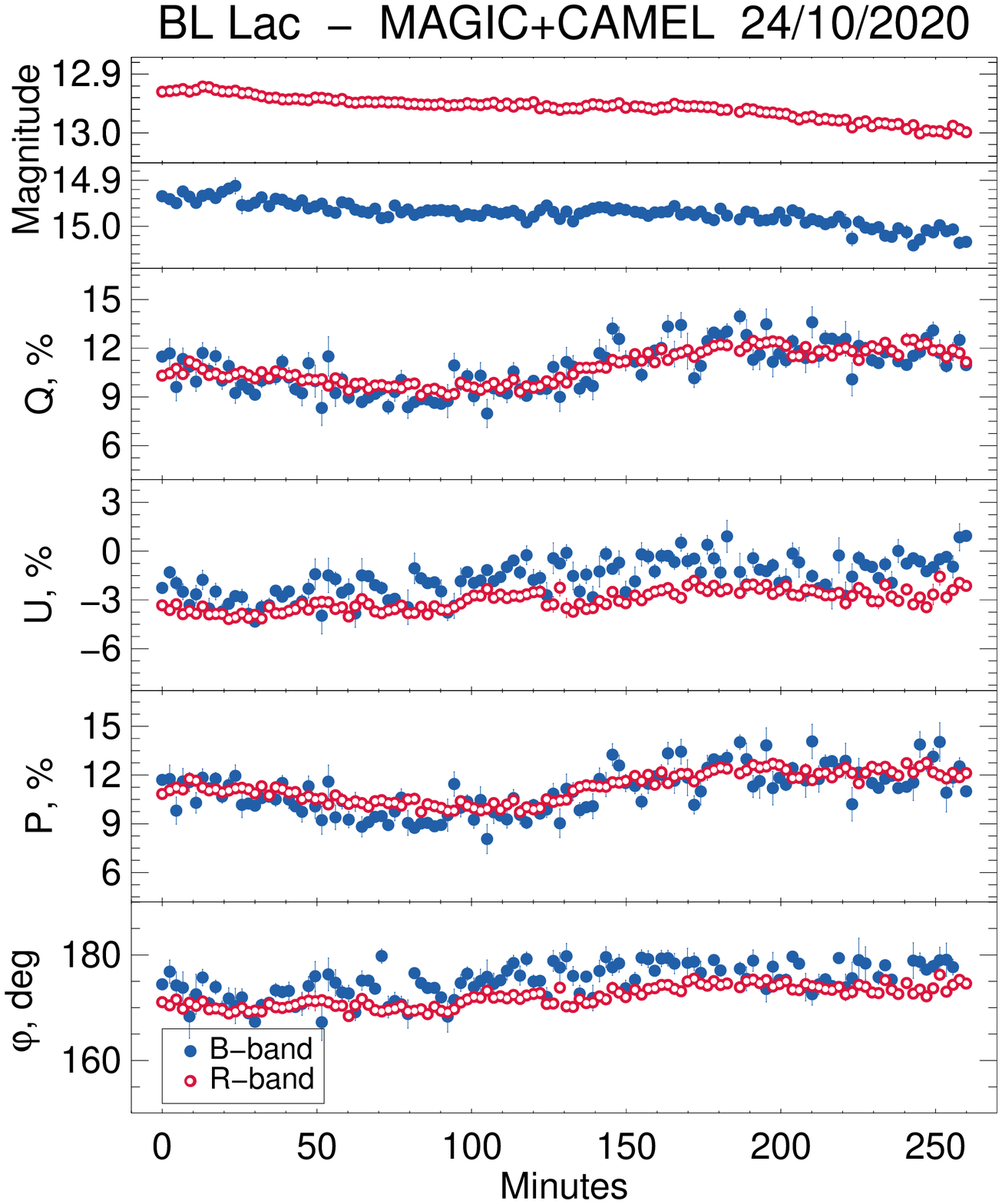}
     \end{subfigure}
     \hfill
     \begin{subfigure}[b]{0.325\textwidth}
         \centering
         \includegraphics[width=\textwidth]{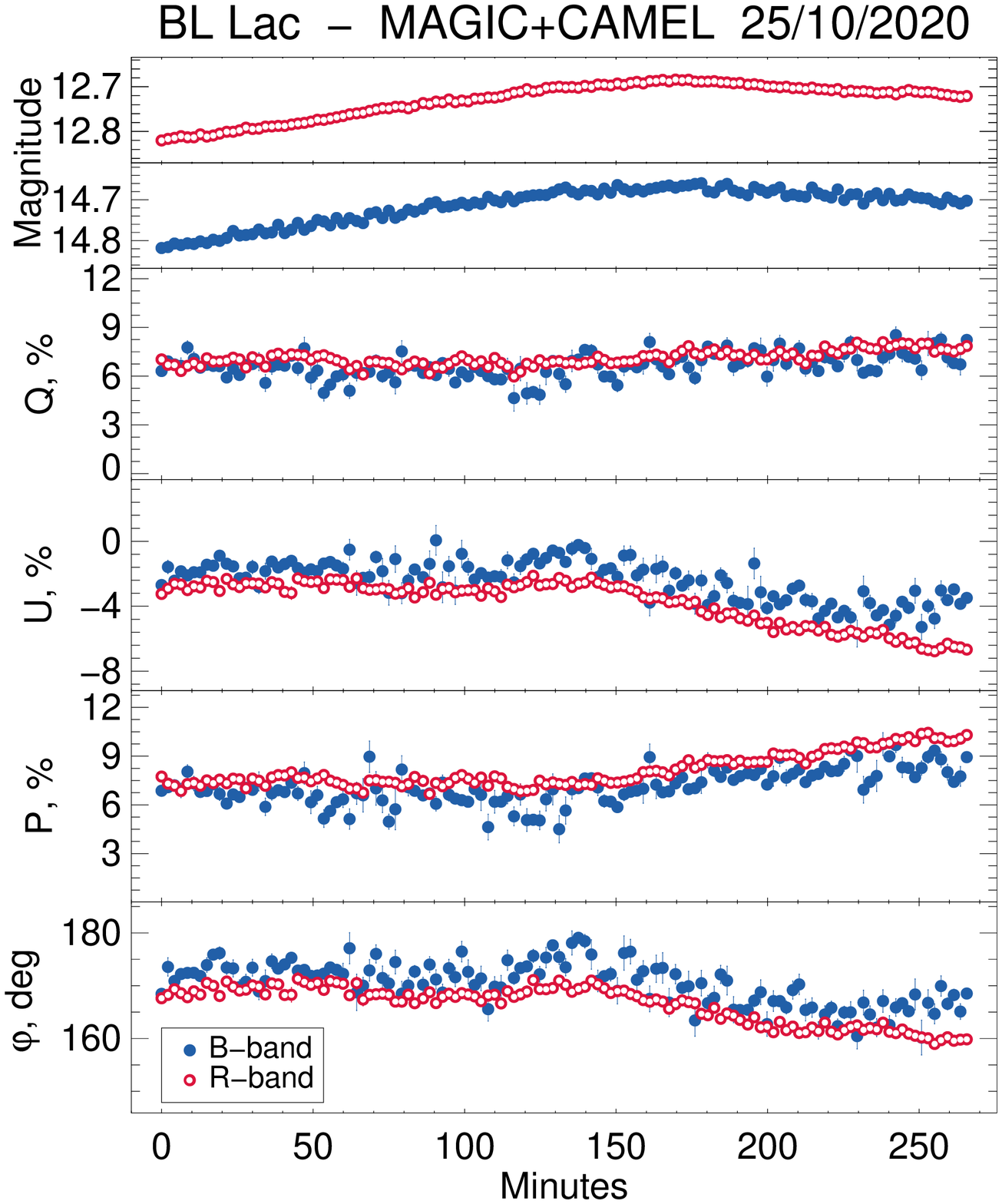}
     \end{subfigure}
     \hfill
     \begin{subfigure}[b]{0.325\textwidth}
         \centering
         \includegraphics[width=\textwidth]{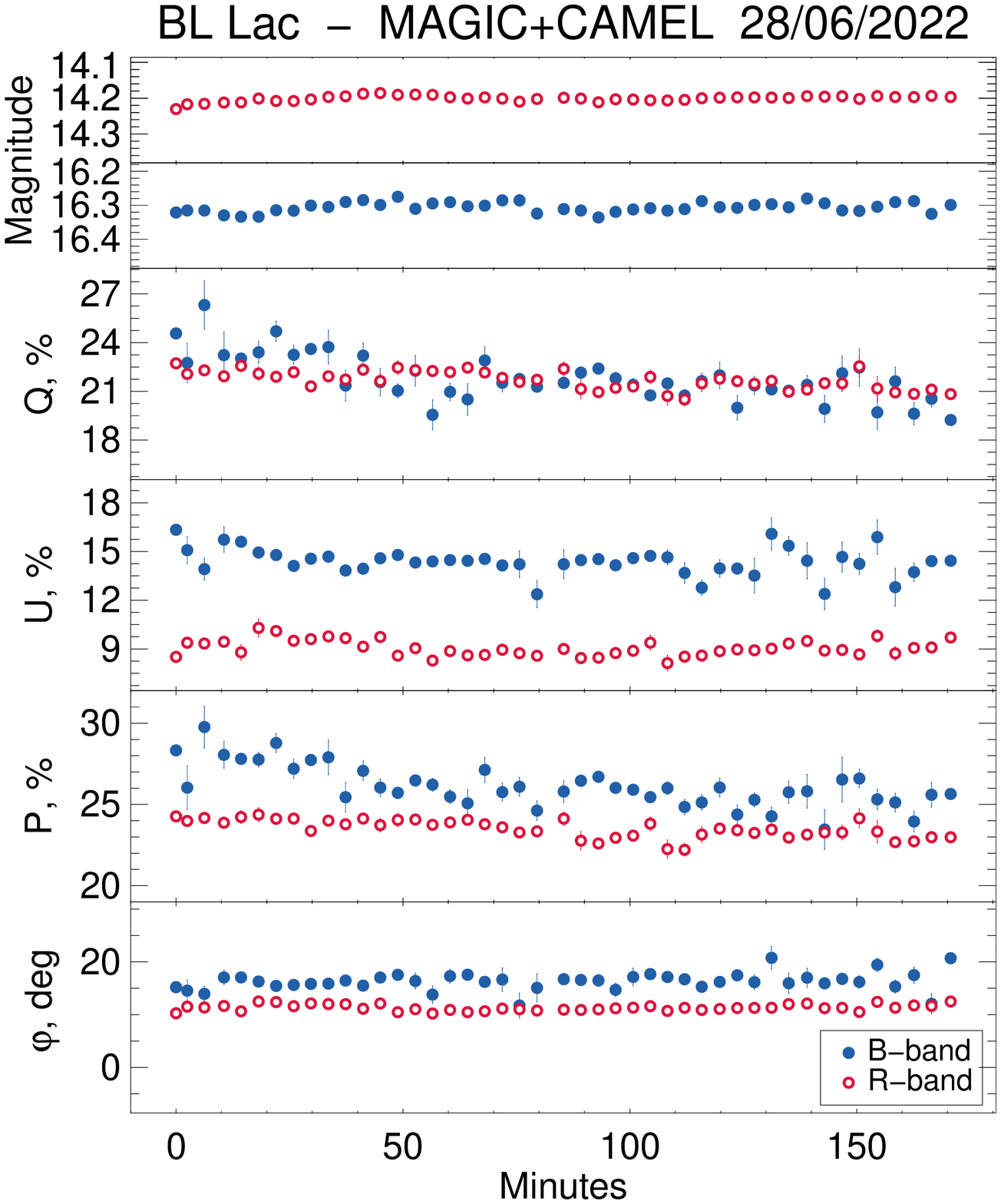}
     \end{subfigure}
    
    \begin{subfigure}[b]{0.325\textwidth}
         \centering
         \includegraphics[width=\textwidth]{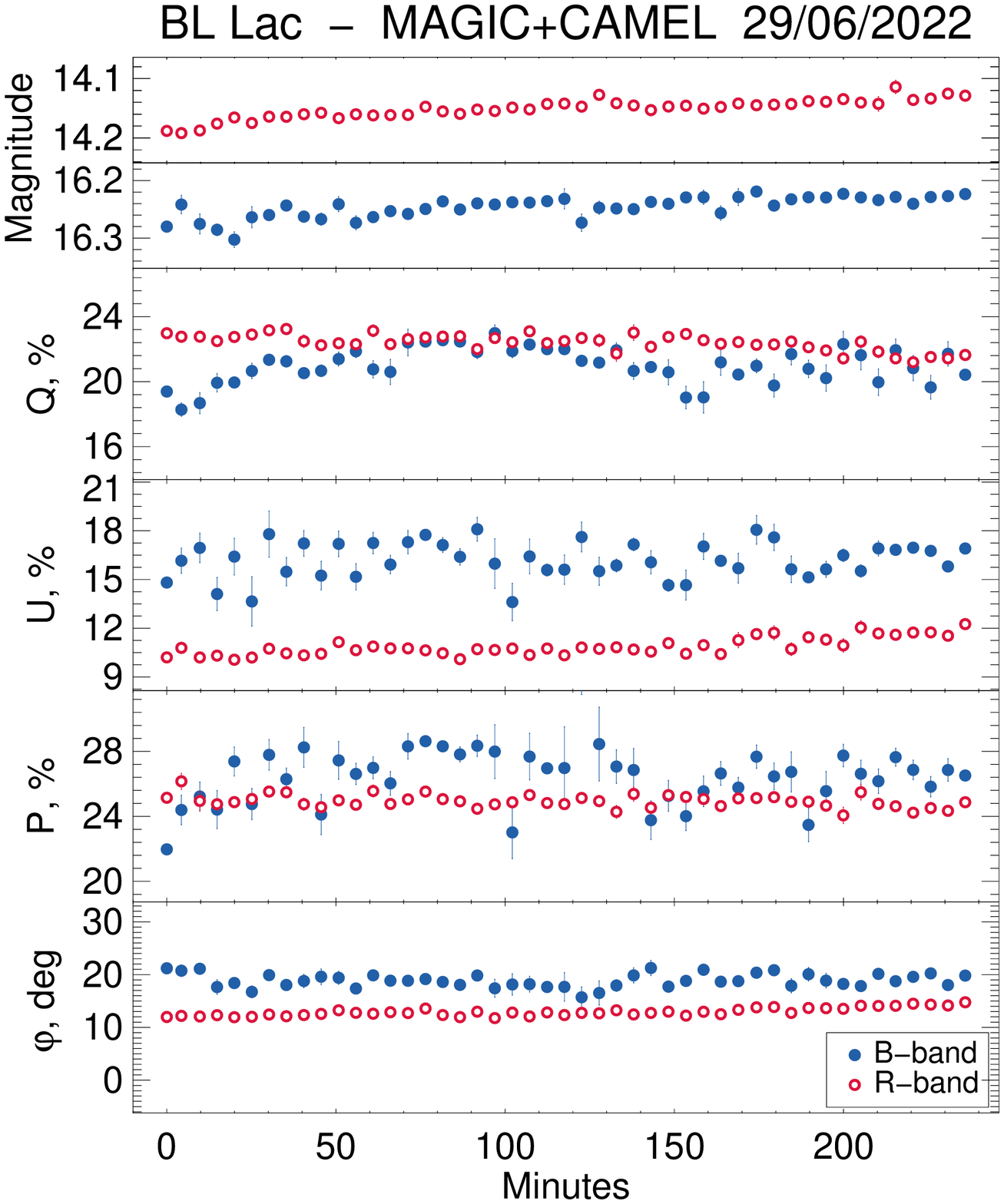}

     \end{subfigure}
     \hfill
     \begin{subfigure}[b]{0.325\textwidth}
         \centering
         \includegraphics[width=\textwidth]{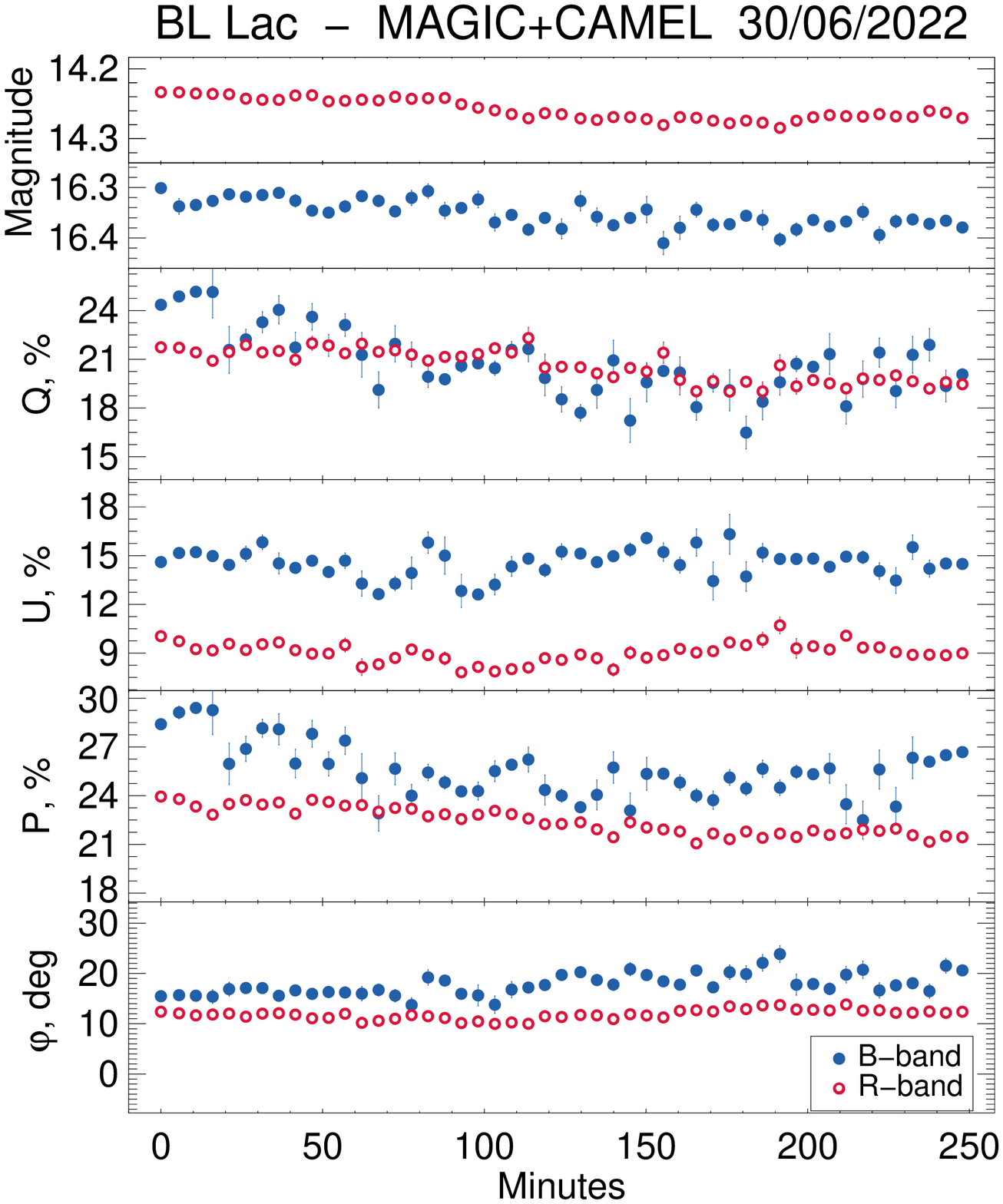}
     \end{subfigure}
     \hfill
     \begin{subfigure}[b]{0.325\textwidth}
         \centering
         \includegraphics[width=\textwidth]{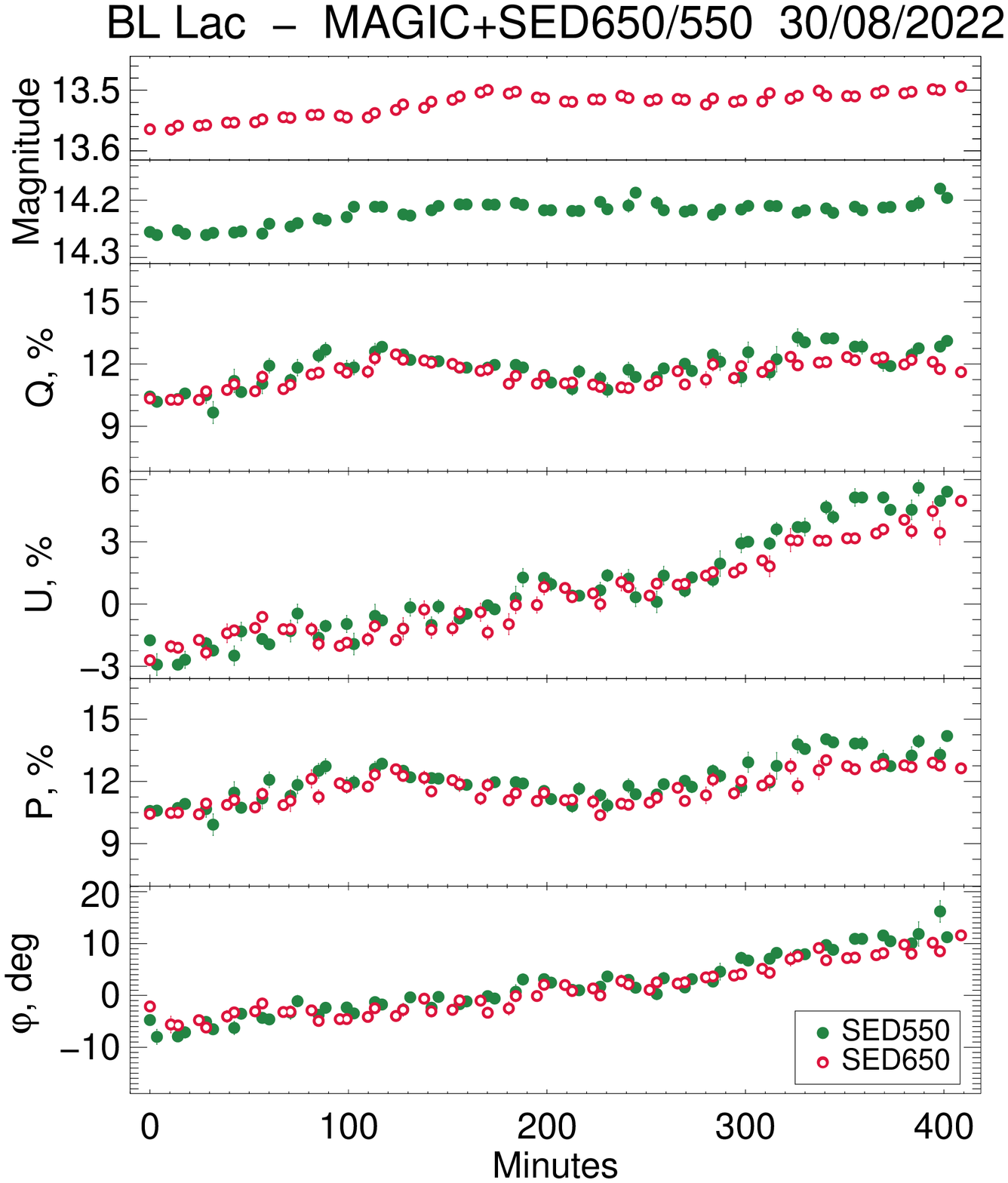}
     \end{subfigure}
        \caption{The light and polarization curves for all epochs when the observations were carried out in two bands. The panels correspond (from top to bottom): magnitude in the "redder"{} band, the magnitude in the "blue"{} band, Stokes parameters $Q$ and $U$ in per cent, the polarization degree $P$ in per cent, the polarization angle $\varphi$ in degrees. The x-axis displays the time in minutes from the beginning of observations in each epoch. The dots colours correspond to the observational bands: filled blue dots denote the $B$ band, filled green dots denote the $V$ band, open red dots denote the $R$ band, and open purple dots denote the $I$ band.}
        \label{LC}
\end{figure*}

In total, 11 epochs of BL Lac observations were obtained in 2020-2022. For 9 epochs, the object was observed in two different colours; 2 more epochs carry additional information in only one colour. For each epoch, the light and polarization curves are shown in Fig. \ref{LC} displaying the variations of magnitude, the Stokes parameters $Q$ and $U$, polarization degree $P$ and angle $\varphi$ with time. To correlate the variability of BL Lac with its state in the context of a long-term trend, for each epoch in each observational band, the average values of brightness, the amplitude of variability, degree and angle of polarization and spectral indices are calculated and given in Table \ref{mean_val}.  

Comparison of the obtained photometric data with open-access monitoring data\footnote{For the analysis, we used open-access data from the American Association of Variable Star Observers (AAVSO) website: \url{https://www.aavso.org /}.} allowed us to identify the states of the blazar activity during the campaign:
\begin{itemize}
    \item[I] -- pre-flare: June-July, 2020;
    \item[II] -- flare: August, 2020;
    \item[III] -- post-flare: October, 2020;
    \item[IV] -- minimum: June, 2022;
    \item[V] -- post-minimum: August, 2022;
\end{itemize}
These states are also presented in Table \ref{mean_val}.

\subsection{Photometric IDV}

As can be seen in Fig. \ref{LC}, for all epochs, even when continuous observations were carried out for only $\sim$1.5 hours, regardless of the object brightness, the magnitude varied by at least 0.05 mag. For a more detailed measurement of the amplitude of the object variability, we use the value $F_{\rm var}$, introduced in the work \citep{Fvar} and often used by other authors to characterize intraday brightness changes:
$$
F_{\rm var} = 100 \times \sqrt{({\rm Max} - {\rm Min})^2 - 2\sigma^2} \ \  (\%), 
$$
where ${\rm Max}$ and ${\rm Min}$ are the maximum and minimum magnitudes of the object during the night, respectively, and $\sigma$ is the average measurement error. The obtained values of the variability amplitudes for each night in each observational band are presented in Table \ref{mean_val}. 
According to the values of $F_{\rm var}$, the greatest amplitude of variability is observed during the flare (II). On the contrary, the object shows weak variability within the night during the period of the deep minimum (IV).

The BL Lac light curves in different epochs of observations demonstrate quasi-sinusoidal changes. By analogy with the study of the variability of the blazar S5 0716+714 \citep{0716,stop}, a wavelet analysis of the BL Lac light curves was carried out. However, in contrast to the case of S5 0716+714, where in two time-separated epochs of observations, the period of oscillation of the brightness was determined with high accuracy as $\sim$1.5 hours, in for BL~Lac it was not possible to uniquely determine the period of variations for any epoch. This behaviour of the object's brightness may indicate either that this period is longer than the duration of observations ranging from 1.5 to 7 hours, or non-stationary processes causing changes in brightness with no harmonic oscillations. {It is important to note that \citet{webb16} also did not reveal any repeatable time-scales of BL Lac IDV during 31 cycles of micro-variability observations.}

Moreover, for all epochs where observations were carried out in two bands, a cross-correlation analysis of the light curves in two colours was carried out. Since the time series are always even, the classical cross-correlation function was used for the analysis. It has been shown that there is no time delay between colours in total light together with $\sim$80-95\% of correlation. {Multiwavelength optical observations presented by \citet{kalita22} stated the same result for their observations in October and November 2020 in g-, r- and i-sdss filters. }

\begin{figure}
    \centering
    \includegraphics[scale=0.4,angle=90]{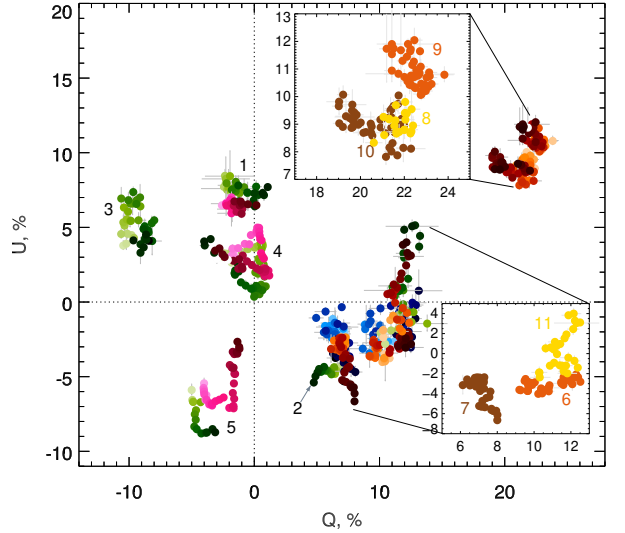}
    \caption{$QU$-diagram for BL Lac. The epochs are indicated by the numbers as in Tables \ref{log_obs} and \ref{mean_val}. The colour change from light to dark corresponds to the time from the beginning of the observations; the data were binned to the uniform cadence of $\sim$8 minutes. The colours correspond to the observational bands: blue is for the $B$ band, green -- for the $V$ band, orange-red -- for the $R$ band and magenta -- for the $I$ band. The inserts with data in only the $R$ band have been added for clarity where the observational data is densely populated.}
    \label{total_qu}
\end{figure}

\subsection{Polarimetric IDV \& \textit{QU}-diagram}

The light curves of the Stokes parameters $Q$ and $U$ and the polarization degree $P$ and angle $\varphi$ are shown in Fig. \ref{LC} using colours matched to the observational bands. The plots show that during all the observed epochs, the polarization changes within the night, and the patterns of these changes differ between the epochs. So, during the flare, we observe violent $P$ and $\varphi$ variations of large amplitude, when the polarization angle was changing dramatically by $\sim$50$^\circ$ in a few hours. Oppositely, during the minimum state, the polarization of BL Lac had a stable angle and a slowly and smoothly varying degree.

\begin{figure*}
    \centering
    \includegraphics[angle=90,scale=0.6]{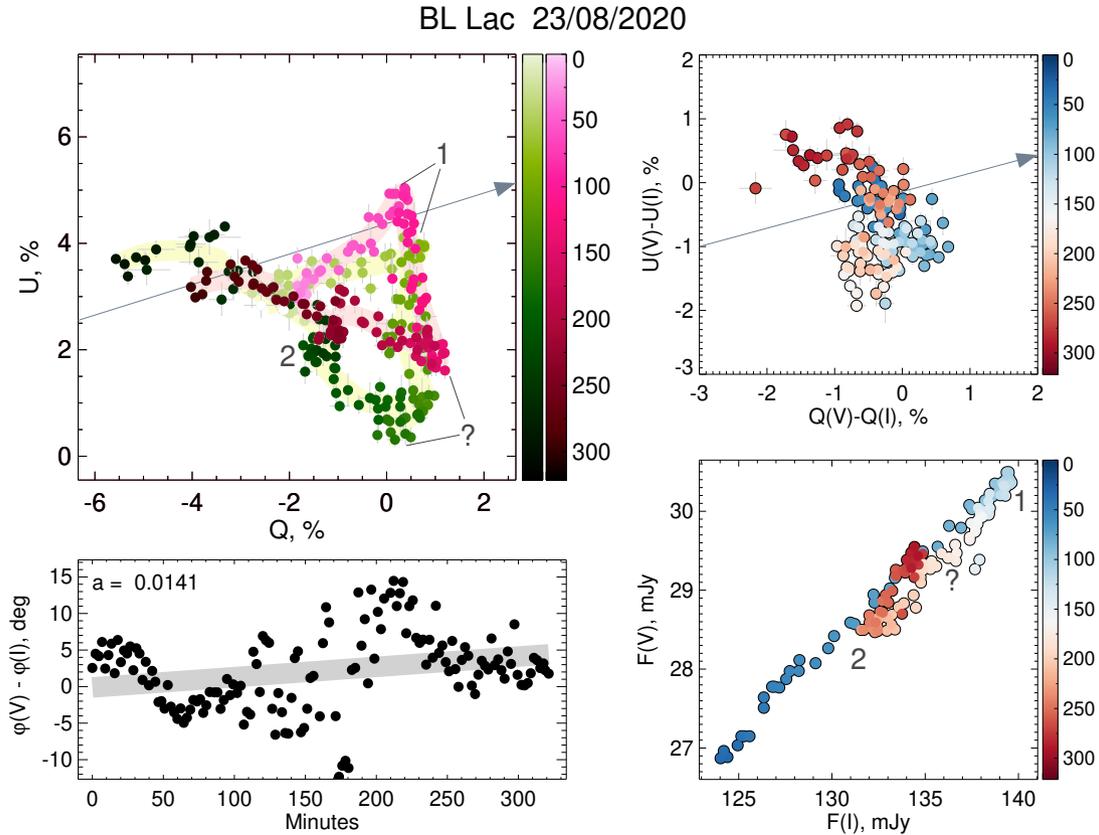}
    \caption{Variations of the polarization and flux of BL Lac during the flare (08/23/2020). \textit{Upper left}: $QU$-diagram in the $V$ (green) and $I$ (magenta) bands. The colour indicates the time from the beginning (light) to the end (dark) of observations. The grey arrow indicates the jet orientation obtained from high-resolution radio data \citep{jet_angle}. \textit{Bottom left}: change of the polarization angle difference $\varphi(V) - \varphi(I)$ in time. \textit{Upper right}: rotation of the differential polarization vector. The colour shows the change in the observation time from blue to red. \textit{Bottom right}: flux-flux diagram, in mJy.}
    \label{fig:flareQU}
\end{figure*}

Furthermore, the mean polarization varies significantly between epochs. It is important to note that during the flare, especially 08/23/2020, the degree of polarization of the object drops to almost zero. This circumstance may indicate possible turbulent processes in plasma (see more in Section \ref{chro} below). The greatest polarization degree $P$ of the order of 25-30\%, which is an outstanding value even for blazars, is observed during the minimum period when the brightness of the object was close to the faintest over the past 8 years ($R \approx 14.3$ mag). {This coincides with the independent measurements of polarization in $R$-band provided in \citep{middei22}.} This is the unexpected result of observations since it is usually assumed that polarization flares should occur simultaneously with the maximum brightness \citep[see e.g.][]{Itoh_2016}. {However, the same anti-correlation was earlier discovered in BL Lac behaviour \citep{gaur14}. }

To demonstrate the polarization vector rotation, the Stokes parameters were depicted on the $QU$-diagram in Fig.~\ref{total_qu}. The epoch numbers are indicated as in Tables \ref{log_obs} and \ref{mean_val}. The colour change from light to dark corresponds to the time from the beginning of the observations; for uniformity, the data is binned over $\sim$8 minutes. The colours correspond to the bands of observations: blue is for the $B$-band, green is for the $V$-band, orange-red is for the $R$-band, and magenta is for the $I$-band. For epochs 8-10, the diagram does not show data in the $B$ band due to the large errors. 

The $QU$-diagram shows that patterns reveal a diverse zoo of trajectories that practically do not repeat. Only in two observational periods -- during the minimum (8-10) and post-flare (6-7) -- the change in the position of the data on the $QU$-plot is small; in other cases, especially during the flare period (4-5), the position on the diagram changes significantly even between two adjacent nights. Qualitatively, this behaviour is typical for a large number of blazars, as can be seen on the $QU$-diagrams in, e.g., \citep{impey}. Some of the longest and most detailed polarimetric observations of BL Lac were carried out in \citep{moore82}. The analysis of the rotation of the polarization vector on the $QU$-plane in that work also showed the randomness of these movements, which is why the rotations were called 'random or drunkard's walk'{}. 

\subsection{Polarimetric IDV in flare}\label{flare}

It is worth considering the IDV of polarization vector in epoch 4 when the maximum brightness of the object was registered over the entire monitoring campaign. On the $QU$-plane in both observational bands (Fig. \ref{fig:flareQU}, upper left), the polarization vector abruptly changes direction several times during the night, forming a "triangular"{} trajectory. Changes in the direction of vector rotation can be considered together with changes in the flux gradient. In the flux-flux diagram (Fig. \ref{fig:flareQU}, bottom right) it can be seen that the radiation intensity of the source first increases in both bands, reaches a maximum, and then also synchronously decreases and begins to rise again. Note that this behaviour is predicted in the case when the variability is determined by a single component with a constant energy distribution \citep{hagen99}. 

Then, if we compare the moments of change in the polarization vector direction on the $QU$-plane with the moments of change in the flux gradient, it turns out that only during event 1 (see Fig. \ref{fig:flareQU}, about 70 minutes after the start of the observations), the rotation of the vector occurs simultaneously with the stop of the increase in the brightness of the source. When the vector rotates the next time (about 160 minutes), this rotation is not reflected in the flux-flux diagram in any way. In turn, the last event on the flux-flux diagram (event 2, about 250 minutes) is represented on the $QU$-diagram by only minor changes in the trajectory.

As previously shown in a number of papers, both long- and short-term changes in the brightness and polarization of blazars can be described within the simple geometric model of the motion of relativistic plasma in a helical magnetic field. Indeed, if we consider the multiparametric model presented in \citep{0716} for a large set of input parameters (e.g. inclination angle, jet opening angle, etc.), then we can achieve similar patterns of polarization change on the $QU$-plane. However, such a model, when compared with real observational data, faces two difficulties. The first is that in the case when polarization variations occur due to geometric effects, the brightness of the source will change due to a change in the observed Doppler factor \citep{but}. However, in this case, the changes in polarization and brightness must occur synchronously, which contradicts observations.

The second important feature found in the observations was that the position of the polarization vector on the $QU$-plane is different in two different colours. Showing the same pattern, the two trajectories are shifted and rotated relative to each other by an angle of $\sim$20$^\circ$. The difference between the two colours also evolves over the observations. Fig. \ref{fig:flareQU}, upper right, shows the motion of the polarization vector on the $QU$-diagram as the difference between two bands of observations, i.e. the polarization vector has coordinates $\{Q(V)-Q(I);U(V)-U(I)\}$. It can be seen that this difference varies over time. Accordingly, the difference in the polarization angles $\varphi(V)-\varphi(I)$ (Fig. \ref{fig:flareQU}, bottom left) also changes. The $\varphi(V)-\varphi(I)$ changes resemble a periodic oscillation, and the extremes correspond to a change in the gradient on a flux-flux diagram.

\section{Polarization chromatism}\label{chro}

\begin{figure}
    \centering
    \includegraphics[scale=0.4,angle=90]{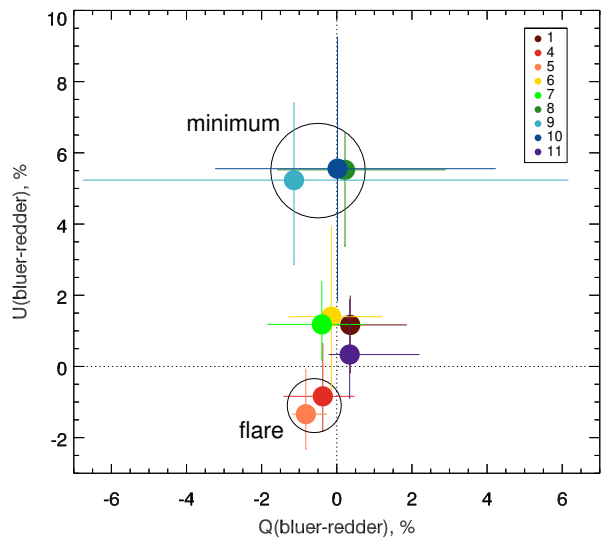}
    \caption{$QU$-diagram for the polarization difference between the "redder"{} and "bluer"{} bands. Different epochs are indicated by different colours given in the legend, the numbers of the epochs correspond to Table \ref{log_obs}. The error bars are indicated formally and correspond to the full spread of points during the epoch.}
    \label{fig:chrom}
\end{figure}

The central result of our observations is that we found that for all epochs there is a dependence of polarization on colour, and the effect is observed both for the degree of polarization (i.e. frequency-dependent polarization degree, FDP) and for the angle (frequency-dependent polarization angle, FDPA). At the same time, the pattern of intraday changes in polarization in different bands of observations is repeated, which is clearly seen by the example of the polarization vector rotation during the flare (Section \ref{flare}) and on similar plots constructed for each of the epochs when observations were carried out in two colours (see Section \ref{A1}). It is worth paying special attention to the fact that the chromatism of polarization varies both from epoch to epoch, and even within the night on the time-scale of hours, which is demonstrated on differential $QU$-diagrams (see Section \ref{A1}). 

To search for the characteristic features of frequency dependence in different brightness states of BL Lac, Fig. \ref{fig:chrom} shows a $QU$-diagram of the polarization difference between the conditionally "red"{} and "blue"{} bands. For the full set of observations, it turned out that, depending on the phase of BL Lac activity, colour polarization is located in different parts of the diagram. In a relatively quiet state (pre-flare, post-flare, post-minimum), the differential polarization is grouped in approximately one place, only slightly different from zero. During the flare, the chromatism changes, and the red polarized component dominates; the differential polarization passes into another quadrant, almost the opposite of the quiet state. {In general, for the most of the epochs the polarization chromatism spreads between -2\% and 2\% both for $Q$ and $U$ parameters. Optical multiwavelength observations of BL Lac conducted by \citet{imazawa22} in 2020 (between 5th and 6th epochs of our monitoring) and 2021 demonstrated the variations of chromatism in the same range. However, according to our data, the exception is the period of the minimum brightness of the object, where }
the chromatism is most sharply manifested. Then not only the maximum polarization of the source is observed (see Fig. \ref{LC}), but also maximum chromatism, not typical for other periods. Unlike the flare period, the blue component dominates the polarization during the minimum state.

Since the blazar radiation shows all the signs of a synchrotron nature, the observed polarization chromatism turns out to be a rather unexpected effect. When we consider the radiation of relativistic electrons in a magnetic field having a power-law energy distribution due to the effective acceleration of any nature (it is usually customary to consider a shock acceleration or magnetic reconnection), the polarization of synchrotron radiation does not depend on the wavelength \citep{synchrotron}. However, over the years of blazars monitoring, chromatism of optical polarization has been repeatedly observed. In the 1980s there was an opinion that the dependence of polarization on wavelength is a rare phenomenon \citep[e.g.][]{saikia88}. However, already \citet{sillanpaa93} found FDP and FDPA are regularly observed in a large number of BL Lac type objects. The FDP of BL Lac itself was shown, for example, in the work \citep{puschell80}, where in the range 0.36-0.69$\mu$m the polarization smoothly changed depending on the wavelength from 5\% to 25\%, reaching a maximum in the blue part of the spectrum. In a number of later works \citep{mead90,sillanpaa93,tommasi01} the FDP of BL Lac was repeatedly confirmed, and in different epochs, both larger polarization in the red spectral range and vice versa were detected. For all the observations available in these papers, differential polarization diagrams similar to Fig. \ref{fig:chrom} were plotted. However, no signs or features of polarization chromatism depending on the brightness of the object were found.

Polarization chromatism is observed for different blazars both in the optical and IR ranges \citep[see][and references therein]{takalo92,takalo94,efimov06}, and, for example, in radio band \citep[see e.g.][]{patk18,krav17}. Moreover, special attention is now focused on the search for the dependence of polarization in the optical and X-ray ranges \citep{liodakis22}. However, if the large difference between the polarization in diverse spectral ranges can be explained by the influence of different physical mechanisms, then the significant dependence of polarization on wavelength and its variability within the optical band is still a subject for discussion. Within the framework of this section, some of the most common mechanisms proposed in the literature to explain the phenomena of FDP and FDPA will be considered.

\subsection{External processes?}

In polarimetric observations, the observed polarization can have not only an internal origin but also a significant contribution from external processes. Among them, the most significant, especially for extragalactic objects, is usually the polarization introduced by ISM. Due to the low galactic latitude ($b\approx -10^\circ$), the contribution of ISM polarization to BL Lac observations is significant, and its accounting is carried out using differential polarimetry using the local standard nearby. As shown above, ISM polarization can be underestimated with this approach, which can introduce bias depending on the wavelength. However, this effect will be small and constant in time on the scale of observations, changing neither between epochs nor within one night of observations. Similarly, we can reject the significant contribution of wavelength-dependent scattering mechanisms (for example, Rayleigh scattering) on the external structures of the AGN, since these processes also do not explain the evolution of chromatism over time.

Also, it should be noted that even during the deep minimum of the object brightness, we observed a high polarization value dominating in the blue part band. This means that the contribution of ISM, scattering processes, as well as depolarization due to the contribution of the host galaxy to the observed optical radiation of BL Lac is negligible. Also, the mechanism of eclipsing the optical jet regions of the blazar by an external cloud belonging to the more distant parts of the jet during the minimum is not confirmed by observations due to "bluer"{} polarization. Moreover, it is important to highlight the potential impact of the accretion disc. \citet{tommasi01}found for BL Lac in a non-extreme phase (13.7 mag in $V$) that the polarization degree increases with the wavelength. It has been suggested that the observed radiation may be significantly influenced by the accretion disc as a thermal component. Similar behaviour was expected for the blazar in any low-brightness state. However, the trend of increasing polarization in the red range with a decrease of the brightness is not observed \citep{mead90,sillanpaa93}, and also completely contradicts the results of BL Lac polarimetry at a minimum state in this work. 
Thus, even during the period of minimum brightness, the observed optical radiation is still dominated by a jet, and not an accretion disc. It could be shown since the polarization models of the accretion disc do not predict a large degree of polarization: in the case of a magnetized disc \citep[see e.g. the model provided by][]{piotrovich}, the observed degree of polarization is $<$2\%, and in the case of Thomson scattering in an accretion disc wind \citep{belob} the maximum degree of polarization in extreme cases can reach 14\%. Also, these models also do not predict polarization variations.

We can also consider the possibility of the influence of Faraday rotation on the optical radiation of a jet propagating in a magnetized medium. Earlier, the Faraday effect has been repeatedly considered in the literature \citep[e.g.][and references therein]{sitko84,diego94}. Due to the difference in the polarization angles in different spectral bands in our observations, it is possible to estimate the value of RM. During the flare, the difference $\Delta\varphi = |\varphi(I) - \varphi(V)| \approx 5^\circ = 0.09$ radians. Then RM = 2.9 $\times$ 10$^{11}$ rad m$^{-2}$. Hence, $<n_eBD> \approx 6\times10^{11}$, where $n_e$ is the electron density in cm$^{-3}$, $B$ is the magnetic field in Gauss, $D$ is the distance in pc. An estimate of the magnetic field strength in an optical jet can be given assuming that the delay between two observational bands is due to synchrotron losses of the high energy electrons \citep{papadakis03}. During the monitoring of BL Lac given in this article, no delay between different colours was detected in any of the epochs for integral or polarized radiation. This gives us an upper limit on the maximum possible delay, equal to the minimum observational cadence, i.e. $\tau$ = 2 minutes = 120 sec. Following the formula from \citep{chiappetti99,papadakis03}:
$$
B\delta^{1/3} \sim 300 \left( \frac{1+z}{\nu_I} \right)^{1/3} \left[ \frac{1-(\nu_I/\nu_V)^{1/2}}{\tau} \right]^{2/3},
$$
where  $\nu_V$ = 0.0055, $\nu_I$ = 0.0034 in units of 10$^{17}$ Hz, we find $B\delta^{1/3} \sim$ 29.5 G, or $B \sim$ 9.5-13.7 G, assuming that 10 $< \delta < 30$ \citep{delta}. The obtained result gives a lower limit assuming given physical assumptions and is more than an order of magnitude higher than the estimates of the magnetic field for BL Lac, obtained earlier usually of the order of 0.5 G.

If we assume that the magnetic field strength in an optical jet is $B\approx 10$ G, then $<n_eD> \approx 6\times10^{10}$. In other words, even if we assume that the size of the region of magnetized matter through which optical radiation passes is $D\sim 1$ kpc, then $n_e\sim 10^7$cm$^{-3}$, which exceeds estimates of the density of gas clouds in NLR ($n^{\rm NLR}_e\sim 10^2 - 10^6$cm$^{-3}$). Based on these calculations, we can confidently conclude that the Faraday effect does not significantly contribute to the observed FDPA. Moreover, during the flare, $\Delta\varphi$ changes sign within the night, which should correspond to a change in the orientation of the magnetic field to the observer. Such significant changes in the orientation of the magnetic field would also be difficult to describe within the framework of modern jet models.

Thus, due to the evolution over time, the main contribution to the variability of polarization and its dependence on wavelength is precisely the result of internal physical processes in the optical jet.

\subsection{Multi-component?}

After detecting the dependence (and the variability of this dependence) of polarization on the wavelength in blazars, the most common interpretation was the model of multi-zone structure radiation. For example, in the work \citep{ballard90}, among others, a simple model is given consisting of two synchrotron components -- polarized and unpolarized.
\citet{holmes84} also proposed a radiation model of two zones with different polarization and spectral indices to explain the polarization characteristics of OJ 287. 

Multi-component models are generally consistent with the commonly accepted consideration of turbulent processes in a jet \citep[see][for a review and references therein]{marscher21}. Simulations show that the presence of turbulent cells in jet plasma can explain the rapid variability of blazars. A variable number of cells increases or decreases the degree of ordering of the magnetic field leading to polarization variability. In plasma, the dominant contribution to radiation passes from one turbulent bubble to another. Due to the different physical conditions, the multi-zone model could explain variations in the chromatism of polarization within the night, especially during the flare. 

The numerical turbulent multi-zone model {\sc TEMZ} was earlier presented in \citep{temz,marscher17}. \citet{marscher21} showed simulation results for the case of a turbulent jet without a shock (see Figure 3 there). There, the greatest variability of polarization chromatism is observed in the case of a 100\% turbulent magnetic field without a regular helical component. Even in this case, the minimum time to change the sign of $dP/d\nu$ is $\sim$0.4 days, i.e. about 10 hours. This is equal to the polarization frequency-dependency should not change significantly during the night. In general, this is the case for most observational epochs (see Fig. \ref{pol_LC}), however, in the case of a flare when turbulent processes are considered dominant, FDP and FDPA change on the scales of $\sim$1 hour. 

It is worth noting that purely turbulent processes in plasma describe qualitatively the observational data quite well. Moreover, the assumption of a significant number of cells in a plasma with different physical characteristics makes it possible to describe any variability by decomposing it into a sufficient number of harmonics. At the same time, the application of such an approach to real observational data faces a number of difficulties. First of all, the cross-correlation analysis of multi-colour observations shows that the variability in different colours is correlated in both total and polarized light. Also, on the $QU$-plane, the rotation of the polarization vector in different colours shows the same patterns up to rotation and shift, regardless of the observation band. This is equivalent to the fact that the radiating region behaves as one zone or, at least, the physical properties of its individual components are interconnected. More importantly, in the results of multi-colour polarization variations given by \citet{marscher21} for the case of plasma without a helical magnetic field, the model shows sharp random jumps in the polarization angle. This behaviour is not confirmed either in this article or by other authors. Polarimetric observations of blazars show that switching and rotation of the polarization angle occur smoothly.

\subsection{Synchrotron and acceleration?}

As mentioned above, the polarization of pure synchrotron emission of a homogeneous source does not depend on the wavelength. Then, to explain the chromatism of polarization, models of the inhomogeneous distribution of the magnetic field and electrons can be considered. In general, for the arbitrary energy distribution of relativistic electrons in an arbitrary magnetic field, the type of polarization dependence on frequency was obtained in \citep{BB82}: $P(\nu) = \Pi(\nu) \cdot (\alpha(\nu) +1)/(\alpha(\nu) +5/3)$, where $\Pi(\nu)$ contains information about the geometry and degree of ordering of the magnetic field, and $\alpha$ is the spectral index. However, such an approach to the description of polarization chromatism, as was indicated in \citep{ballard90}, has no direct relation to the physical model. Alternatively, \citet{ballard90} suggests, like \citet{bjo85}, considering synchrotron radiation of cut-off electron distribution. In that work, Figure 9 shows that in the presence of cut-off frequency, the dependence of polarization on frequency has an asymmetric quasi-bell-shaped distribution with a maximum degree of polarization at the cut-off frequency. In that work, for a large data set of multi-wave polarization of a sample of blazars, the data were approximated using this model, with the cut-off frequency as a free parameter, which made it possible to describe observational data with any sign of $dP/d\nu$. Thus, FDP can be described qualitatively and quantitatively by the cut-off electrons synchrotron radiation model. However, this approach excludes the existence of FDPA, which rises problems with describing BL Lac data.

The assumption described by \citet{ballard90} is physically justified, since the cut-off can be both a consequence of the presence of the maximum frequency of electron emission (when energy losses are equal to the acceleration energy gain) and, e.g., a passing shock wave. In general, consideration of changes in the physical characteristics of plasma due to the presence of shock processes in the jet \citep{shock} is the most popular approach today. The shock wave as an effective internal mechanism of electron acceleration in jets is capable of simultaneously causing rapid flux variability in a wide range of wavelengths and significant changes in the electron energy distribution. In particular, the energy-stratified electron distribution caused by the impact should generate strongly chromatic polarized radiation, which is observed when comparing optical and X-ray polarimetric data \citep{liodakis22}.

{As shown by \citet{imazawa22}, shock-in-jet model alone could not explain all observed features, e.g. higher polarization degree in optical band than in near-IR. \citet{webb17,webb21} provided polarimetric observations of the sample of blazars in $VRI$ and $H$ bands aimed at testing of the two-zones shock in jet model \citep{krm}. This model is based on the shock excitement of turbulent cells, cooling then through synchrotron radiation. This approach demonstrated excellent agreement with photometric IDV \citep{webb16,webb21}. However, polarimetric IDV as a critical test on the original failed to show the predicted polarization behaviour, and the archival data \citep{sasada08,bhata15} was too dense to draw any conclusion.   }
Using the TEMZ model for the case of a standing shock \citet{marscher21} calculated the polarization in the optical bands $B$, $R$, $I$. {These predictions are more suitable not only for the results obtained in this paper but for all optical campaigns of BL Lac taken during 2020-2022.} Unlike the model without a shock, here, even in the absence of the helical component of the magnetic field, the change in the polarization angle occurs smoothly, which is generally consistent with observations. FDP and FDPA are also observed regardless of the configuration of the magnetic field. Although TEMZ results qualitatively agree well with the BL Lac polarimetry data, they still cannot quantify them. First of all, due to the presence of a shock, the chromatism of the optical polarization changes over a $\sim$2 times longer time interval, which eliminates the variability of FDP on the scale of hours. Moreover, the maximum chromatism between $B$ and $R$ bands is on the order of $\sim$2\% of the degree of polarization and does not exceed $\sim$10$^\circ$, which also does not agree with the data.

{
Magnetic reconnection \citep{mizuno11,sironi15,petr16} is considered to be an alternative acceleration mechanism possible to explain the violent variability of the jet emission. Comparing with the shock wave models, magnetic reconnection events also provide variable polarization in all electromagnetic range yet the variability time-scales are shorter and of the order of an hour or several hours \citep{zhang22}, which coincides with known polarimetric IDV measurements. Moreover, according to the present simulations \citep{zhang20}, magnetic reconnection leads to the polarization chromatism of a detectable amplitude of at least several per cents between optical and near-IR bands. Qualitatively this approach agrees with the observational data. Further, much more exact approximation of available polarimetric data via reconnection simulations should be worked out. 
}

\section{Conclusion}\label{con}

A significant number of epochs of optical polarimetric observations of BL Lac obtained during 2020-2022 allowed not only determining the rapid changes in the brightness and polarization of BL Lac but also correlating these changes with long-term variability. 

1. In any state of activity, BL Lac demonstrates photometric IDV, even when the object is in the minimum phase. Cross-correlation analysis showed no time delay between the different colours in which observations were carried out quasi-simultaneously during each epoch. This allows one to put the upper limit on the delay equal to the minimum cadence time of 2 min. A wavelet analysis of the observational data was also carried out, but periodic changes in the integral and polarized fluxes could not be identified. 

2. Polarimetric IDV is observed in any state of BL Lac activity. The $QU$-plane showed that the polarization state does not have a priority location or a phase-dependent position on the diagram. During the period of activity, both the degree and the angle of polarization change, and the polarization degree is as low as $\sim$3\%. During the minimum state, there are practically no changes in the polarization angle, the polarization degree varies slightly and smoothly, but is of $\sim$25-30\%. Qualitatively, this polarization behaviour is consistent with the dominance of turbulent processes in the plasma during the flare and, conversely, a higher degree of organization of the magnetic field during the minimum brightness state of the blazar.

3. All epochs of BL Lac observations are characterized by the presence of chromatism of optical polarization. The variability of this chromatism between epochs and inside the night cannot be explained by any external mechanisms or the influence of components outside the jet. Thus, the variable chromatism of polarization indicates internal mechanisms in the optical jet capable of significantly changing the physical parameters of the emitting plasma on scales up to 1.5 hours. Many models of inhomogeneous (broken) electron distribution, {including shock-in-jet and magnetic reconnection models}, predict the presence of chromatism of optical polarization. However, none of the models discussed above can fully describe the obtained observational data quantitatively. The main challenges for the models are (i) the BL Lac flare when the sign of $dP/d\nu$ changed on the scales of several hours during the night, and the polarization degree was quite low, and (ii) the period of a deep minimum of the object brightness, when the jet still dominates, and the polarization degree turns out to be maximal for the entire set of observations and the most wavelength-dependent.

\section*{Acknowledgements}

The work was performed as part of the SAO RAS government contract approved by the Ministry of Science and Higher Education of the Russian Federation. Observations with the SAO RAS telescopes are supported by the Ministry of Science and Higher Education of the Russian Federation.

\section*{Data Availability}

The reduced data underlying this article is available upon the request to the corresponding author 1 yr after the publication of this paper.



\bibliographystyle{mnras}
\bibliography{biblio} 




\appendix

\section{BL Lac light curves of polarization and brightness}\label{A1}

\begin{figure*}
     \centering
     \begin{subfigure}[b]{0.48\textwidth}
         \centering
         \includegraphics[width=\textwidth]{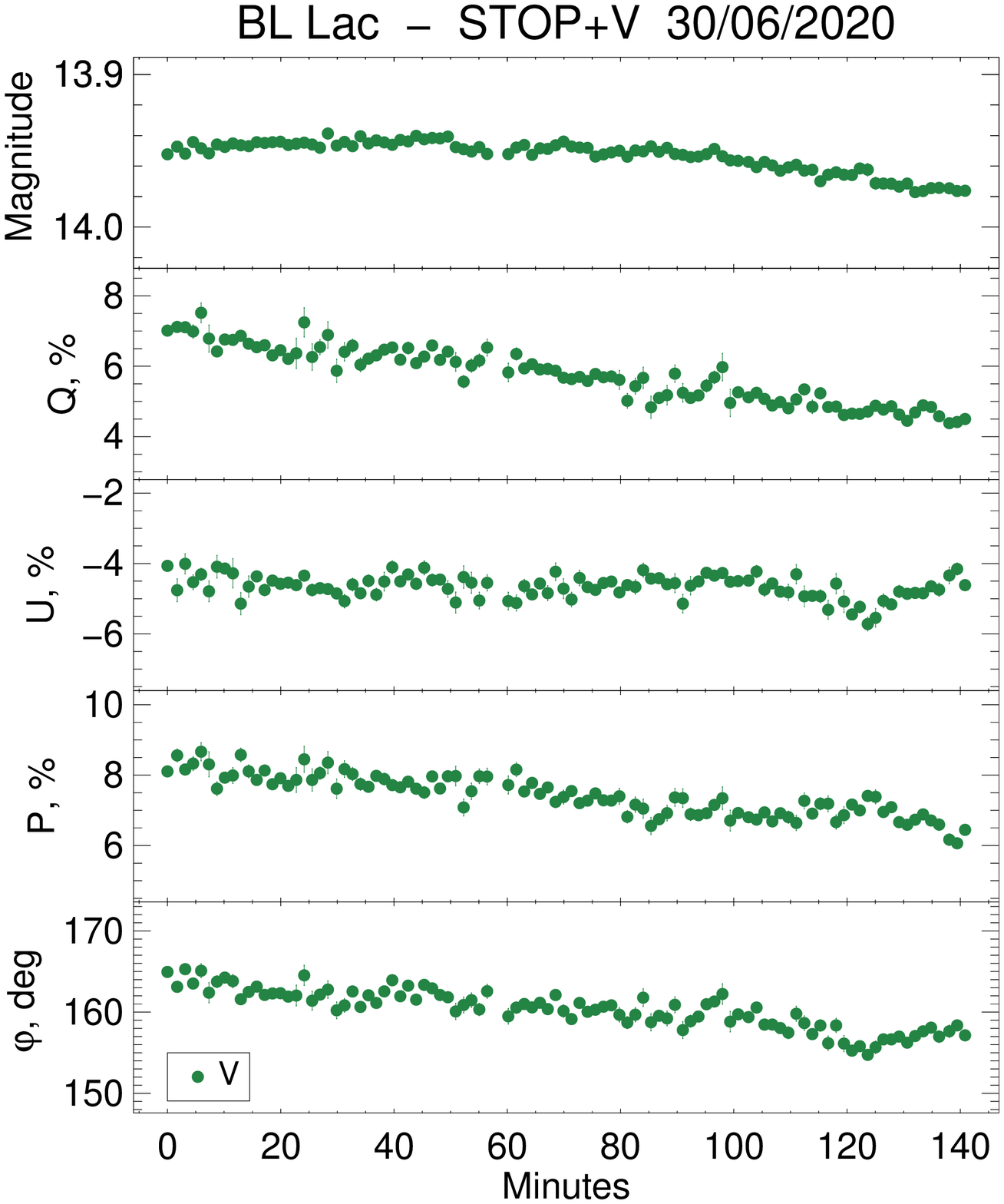}
     \end{subfigure}
     \hfill
     \begin{subfigure}[b]{0.48\textwidth}
         \centering
         \includegraphics[width=\textwidth]{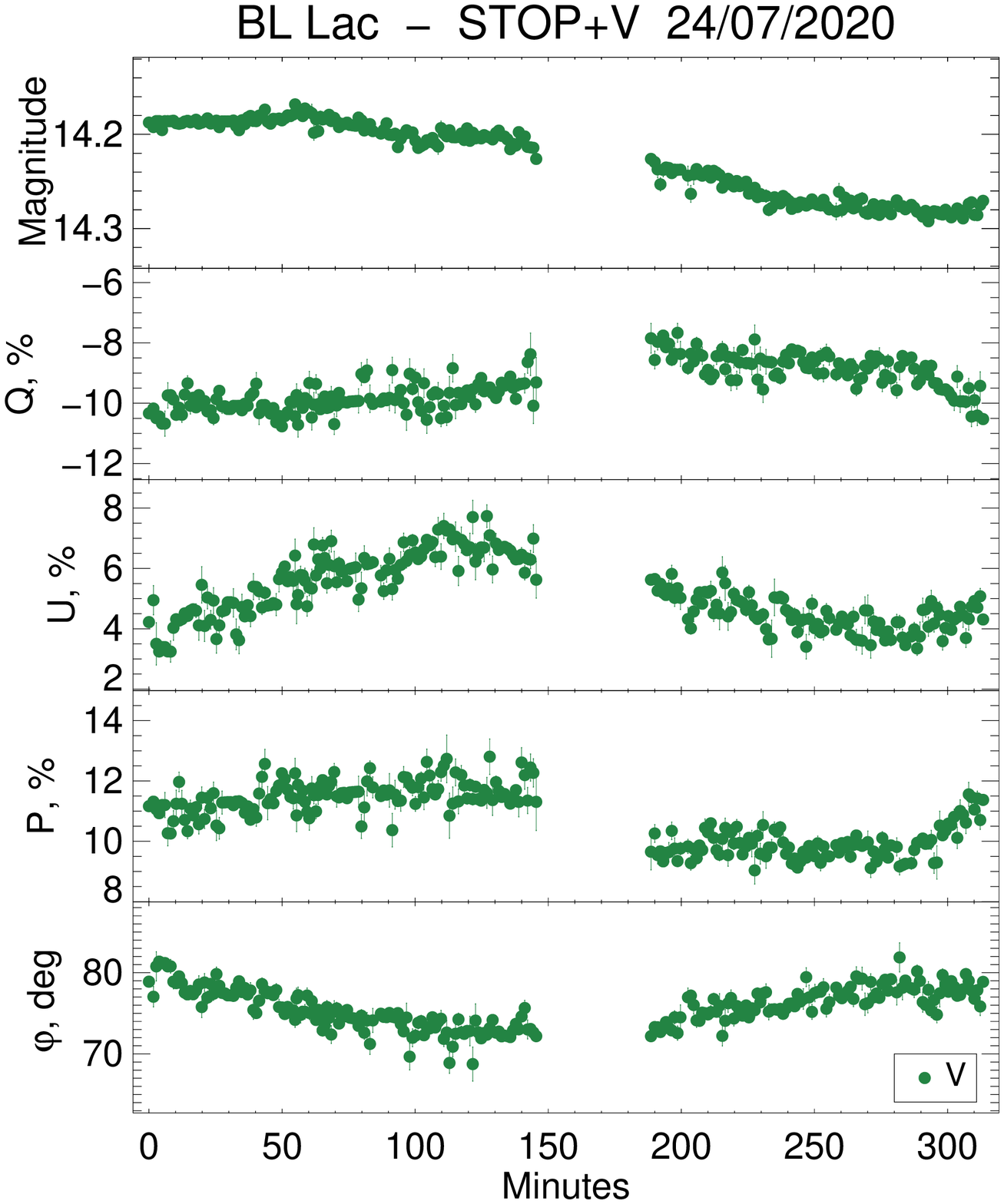}
     \end{subfigure}
        \caption{The light and polarization curves for the epochs in which observations were carried out in one band. The panels correspond (from top to bottom) to: the magnitude in the $V$ band, the Stokes parameter $Q$ and $U$ in per cent, the degree of polarization $P$ in per cent, and the polarization angle $\varphi$ in degrees. The x-axis displays the time in minutes from the beginning of observations in each epoch. }
        \label{ap1}
\end{figure*}

\begin{figure}
    \centering
    \includegraphics[angle=90,scale=0.37]{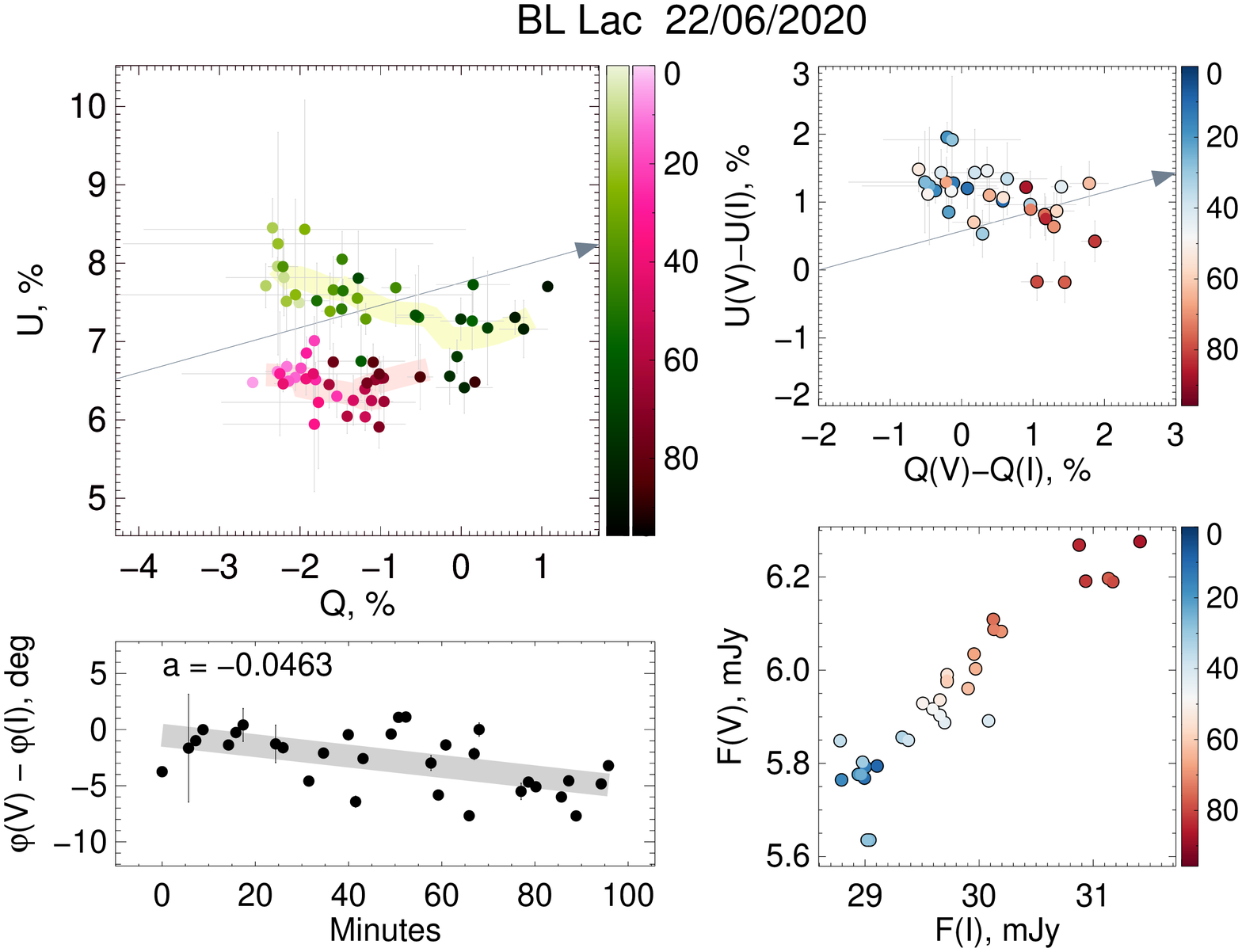}
    \caption{Variations of the polarization and flux of BL Lac at 22/06/2020. \textit{Upper left}: $QU$-diagram in the $V$ (green) and $I$ (magenta) bands. The colour indicates the time from the beginning (light) to the end (dark) of observations. The grey arrow indicates the jet orientation. \textit{Bottom left}: change of the polarization angle difference $\varphi(V) - \varphi(I)$ in time. \textit{Upper right}: rotation of the differential polarization vector. The colour shows the change in the observation time from blue to red. \textit{Bottom right}: flux-flux diagram, in mJy.}
    \label{A2}
\end{figure}

\begin{figure}
    \centering
    \includegraphics[angle=90,scale=0.37]{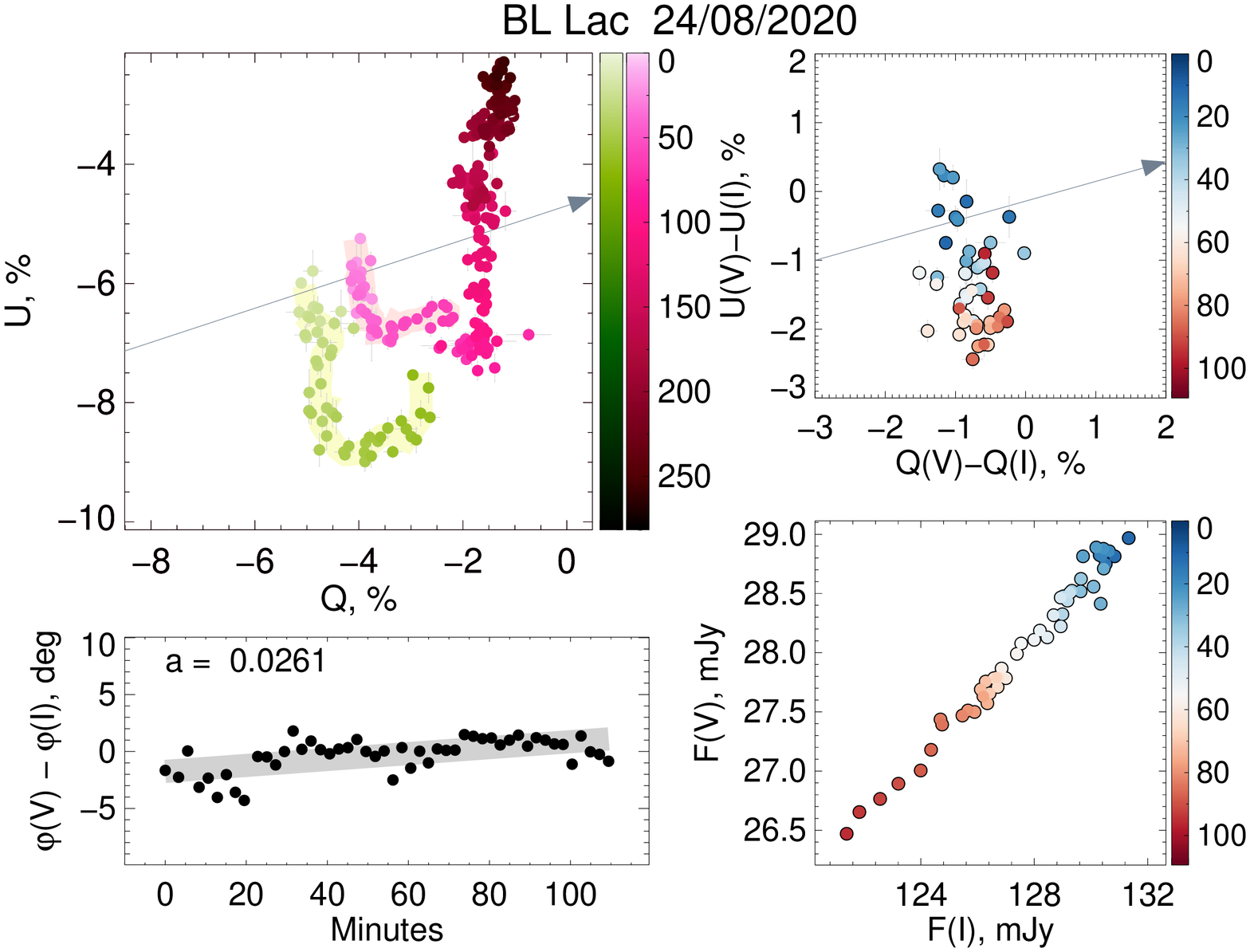}
   \caption{Similarly to Fig. \ref{A2} for the $V$ (green) and $I$ (magenta) bands in the epoch 24/08/2020.}
    \label{A3}
\end{figure}

\begin{figure}
    \centering
    \includegraphics[angle=90,scale=0.33]{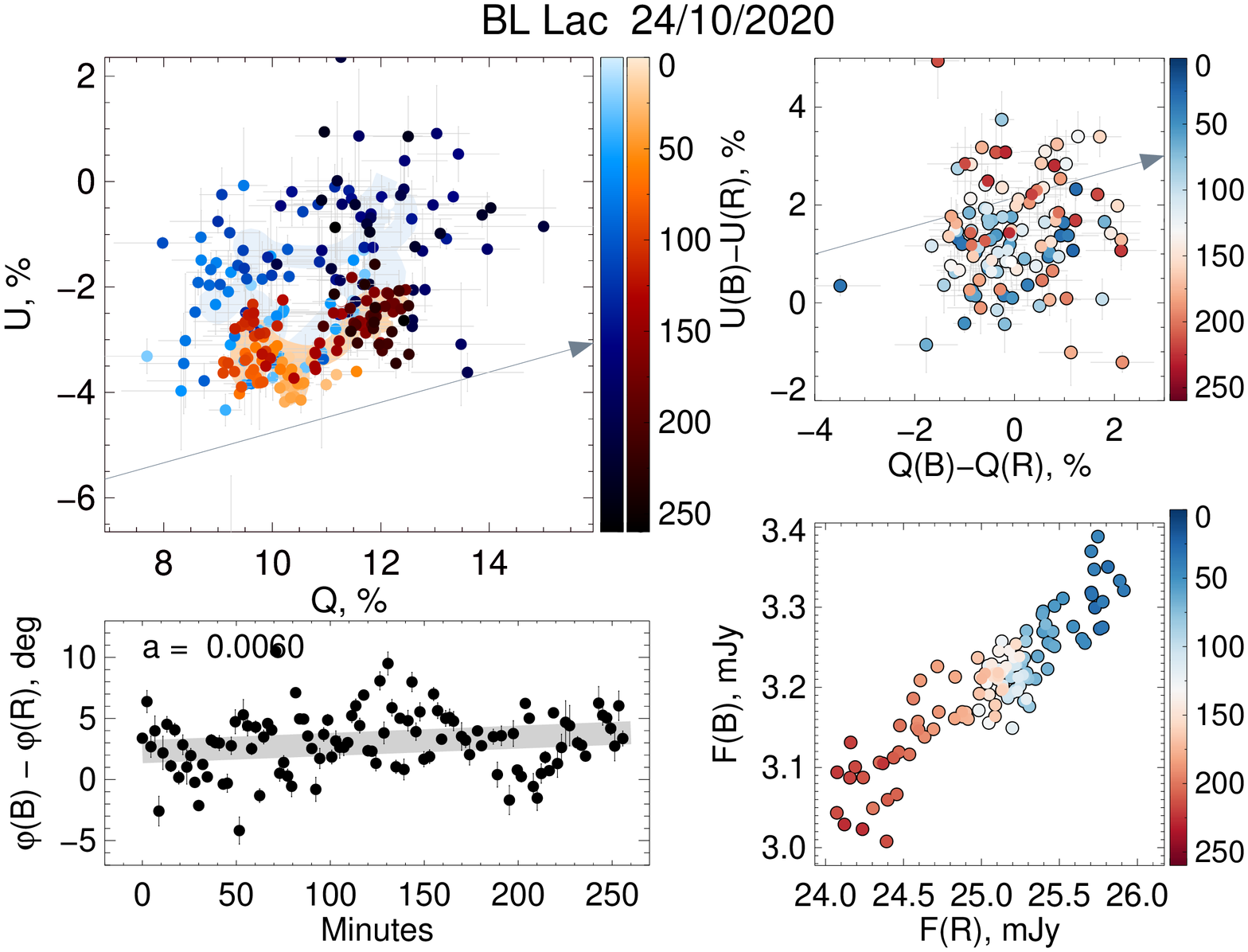}
    \caption{Similarly to Fig. \ref{A2} for the $B$ (blue) and $R$ (orange-red) bands in the epoch 24/10/2020.}
        \label{A4}
\end{figure}

\begin{figure}
    \centering
    \includegraphics[angle=90,scale=0.33]{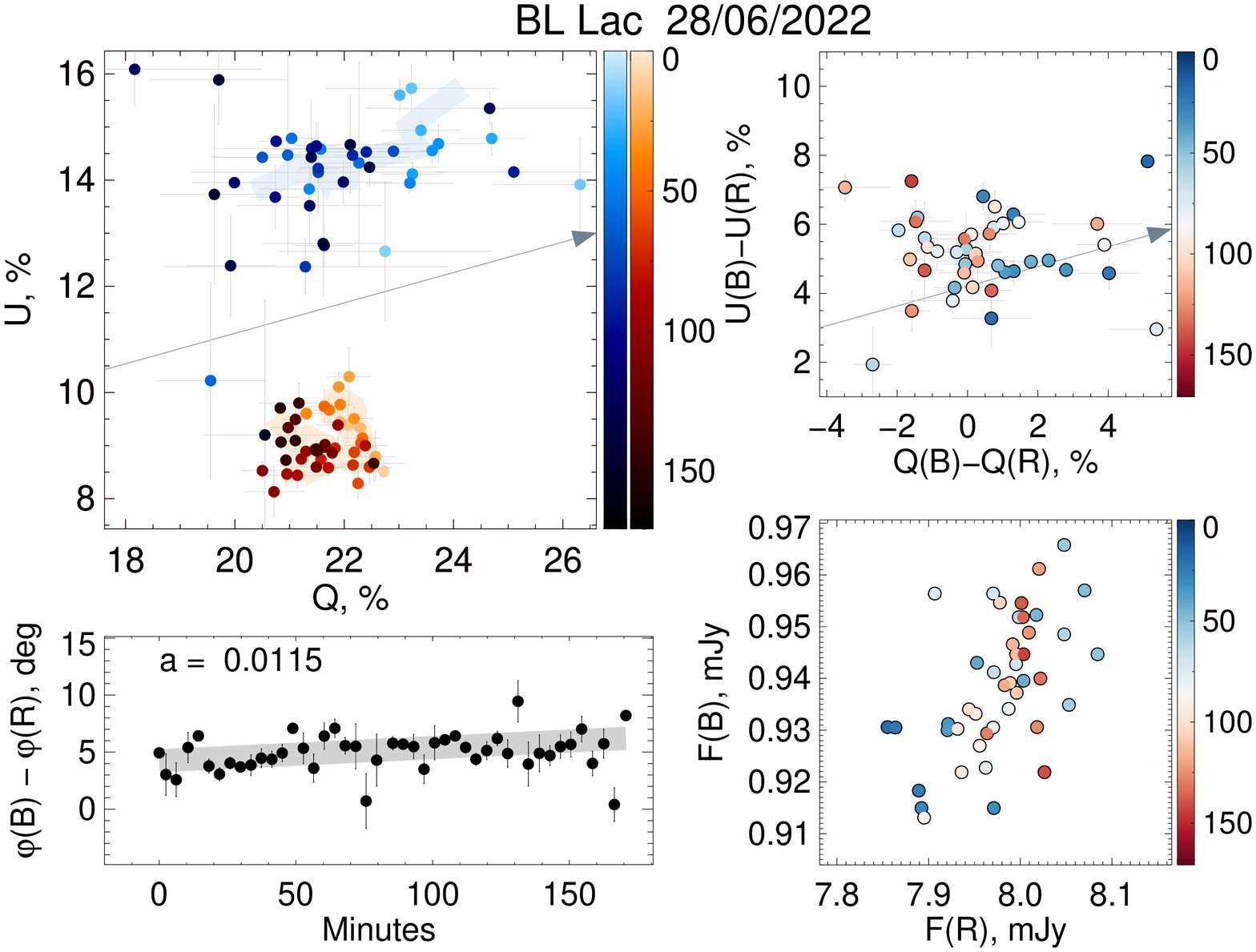}
    \caption{Similarly to Fig. \ref{A2} for the $B$ (blue) and $R$ (orange-red) bands in the epoch 28/06/2022.}
        \label{A5}
\end{figure}

\begin{figure}
    \centering
    \includegraphics[angle=90,scale=0.33]{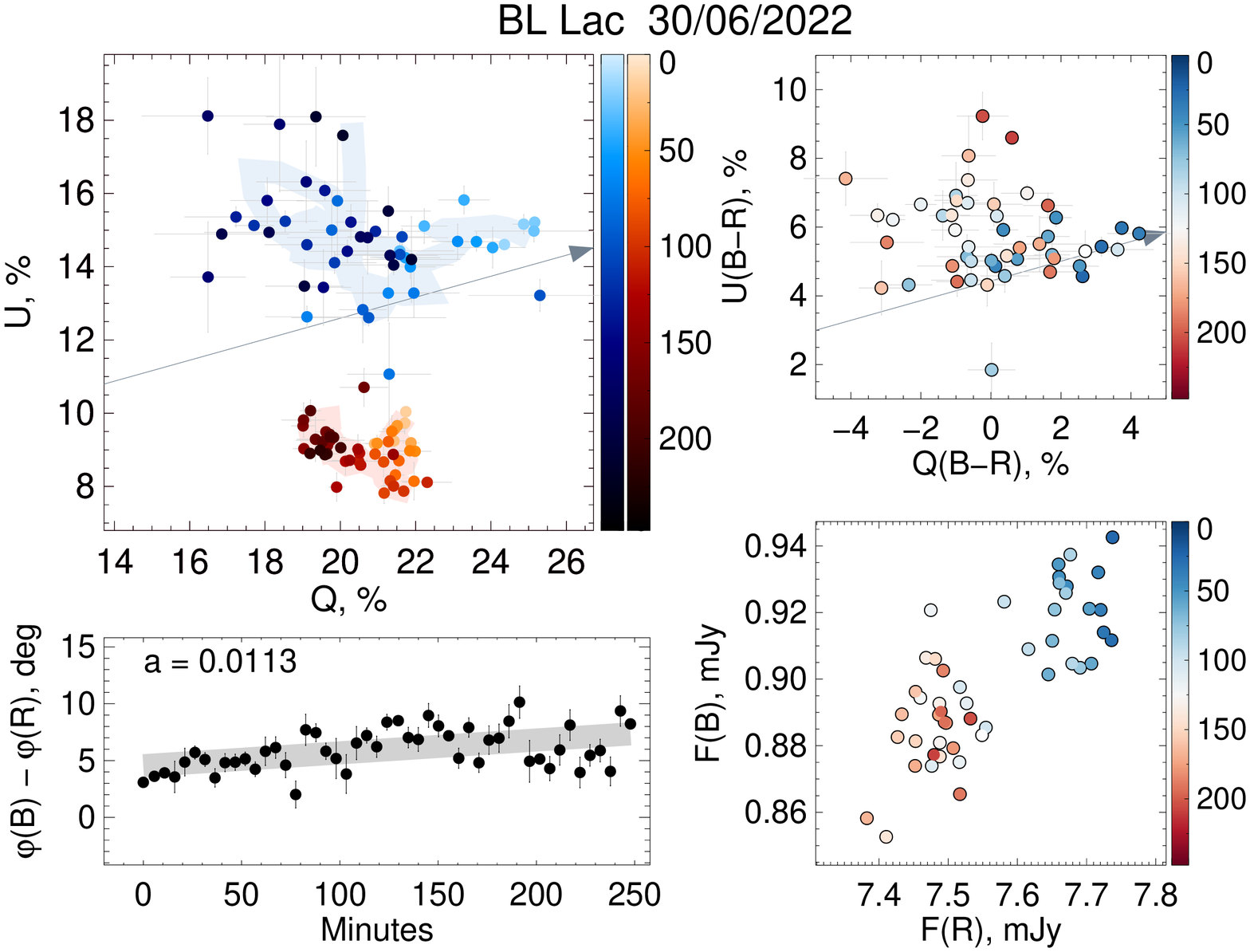}
    \caption{Similarly to Fig. \ref{A2} for the $B$ (blue) and $R$ (orange-red) bands in the epoch 30/06/2022.}
        \label{A6}
\end{figure}

\begin{figure}
    \centering
    \includegraphics[angle=90,scale=0.33]{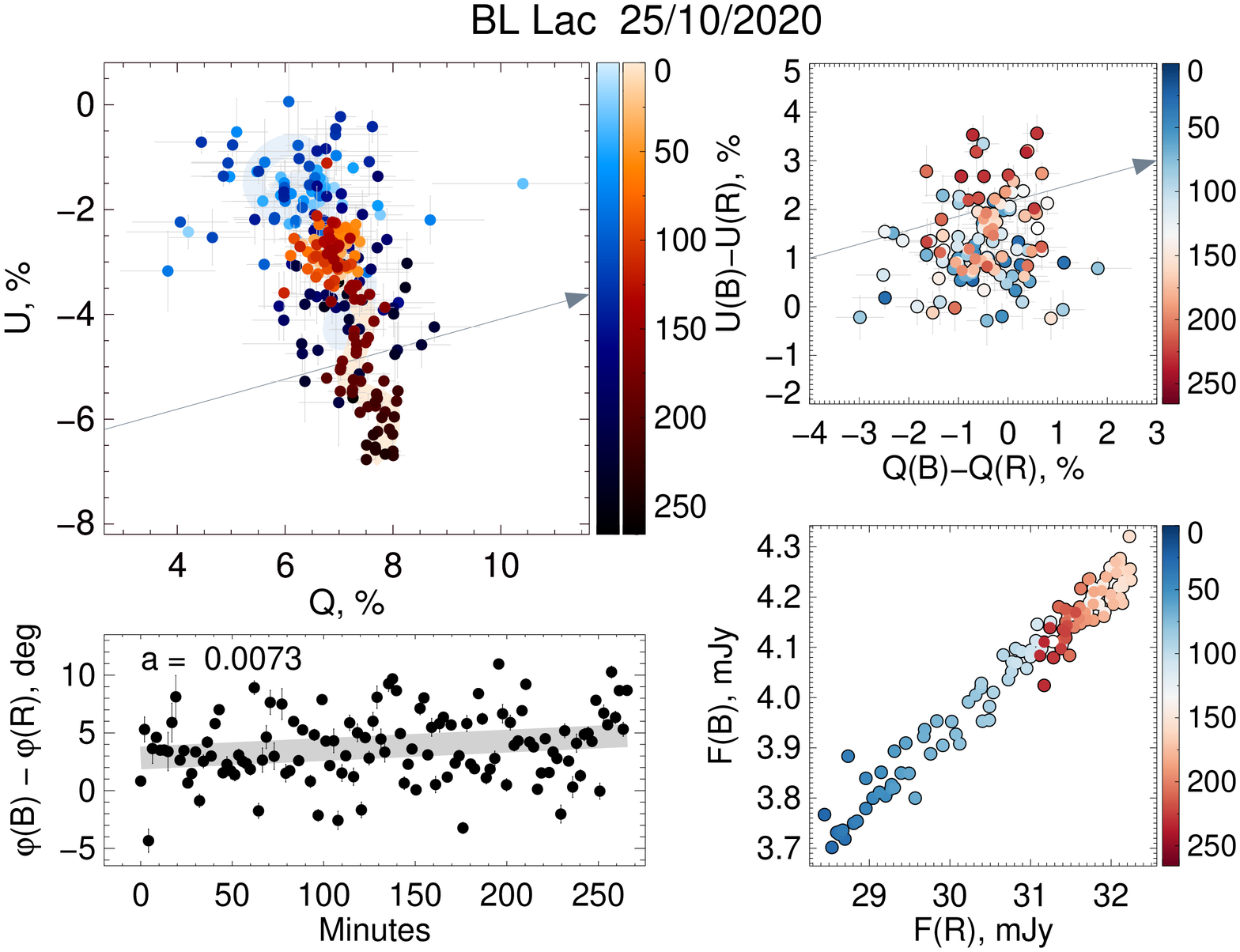}
    \caption{Similarly to Fig. \ref{A2} for the $B$ (blue) and $R$ (orange-red) bands in the epoch 25/10/2020.}
        \label{A7}
\end{figure}

\begin{figure}
    \centering
    \includegraphics[angle=90,scale=0.33]{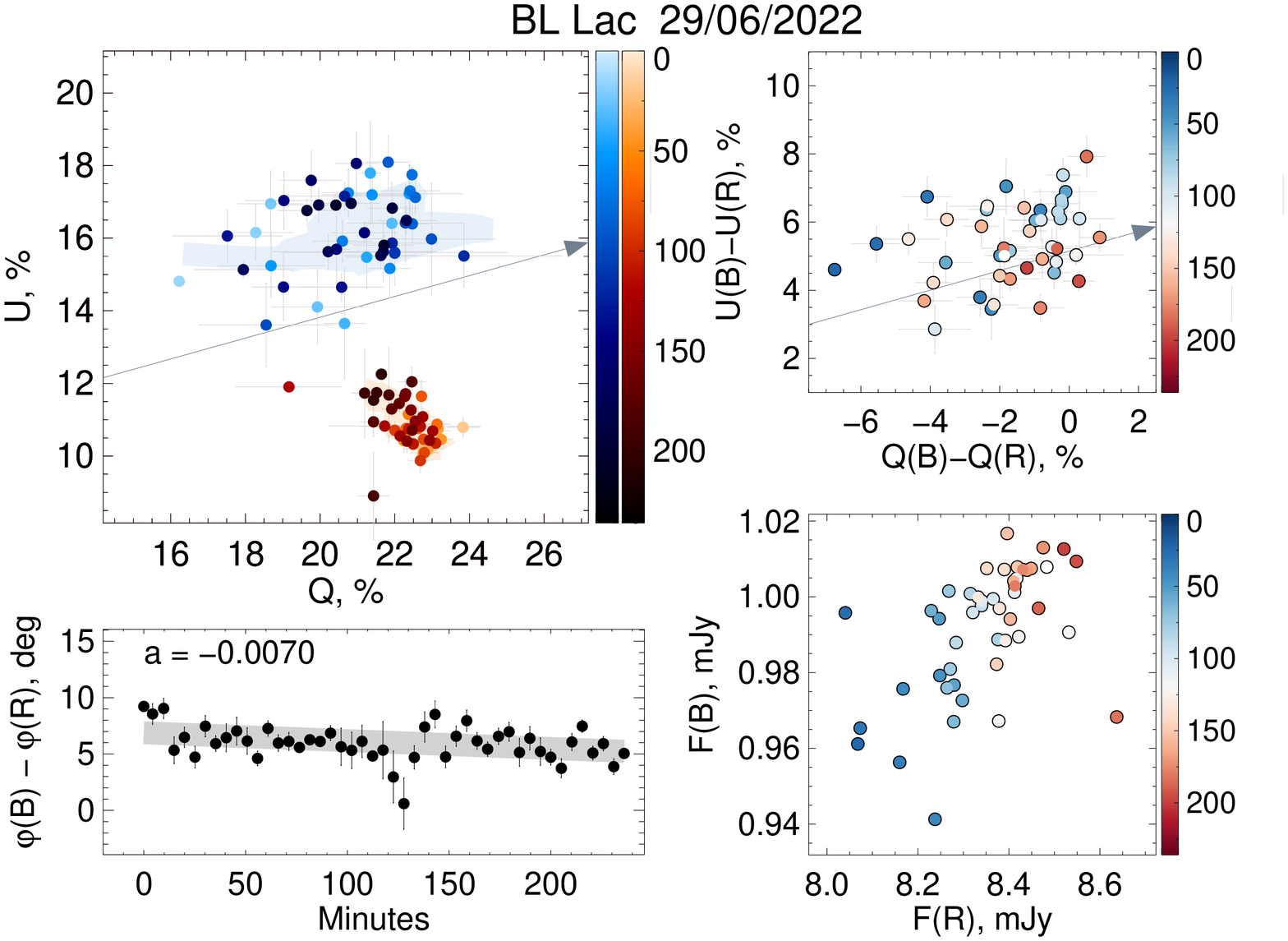}
    \caption{Similarly to Fig. \ref{A2} for the $B$ (blue) and $R$ (orange-red) bands in the epoch 29/06/2022.}
        \label{A8}
\end{figure}

\begin{figure}
    \centering
    \includegraphics[angle=90,scale=0.33]{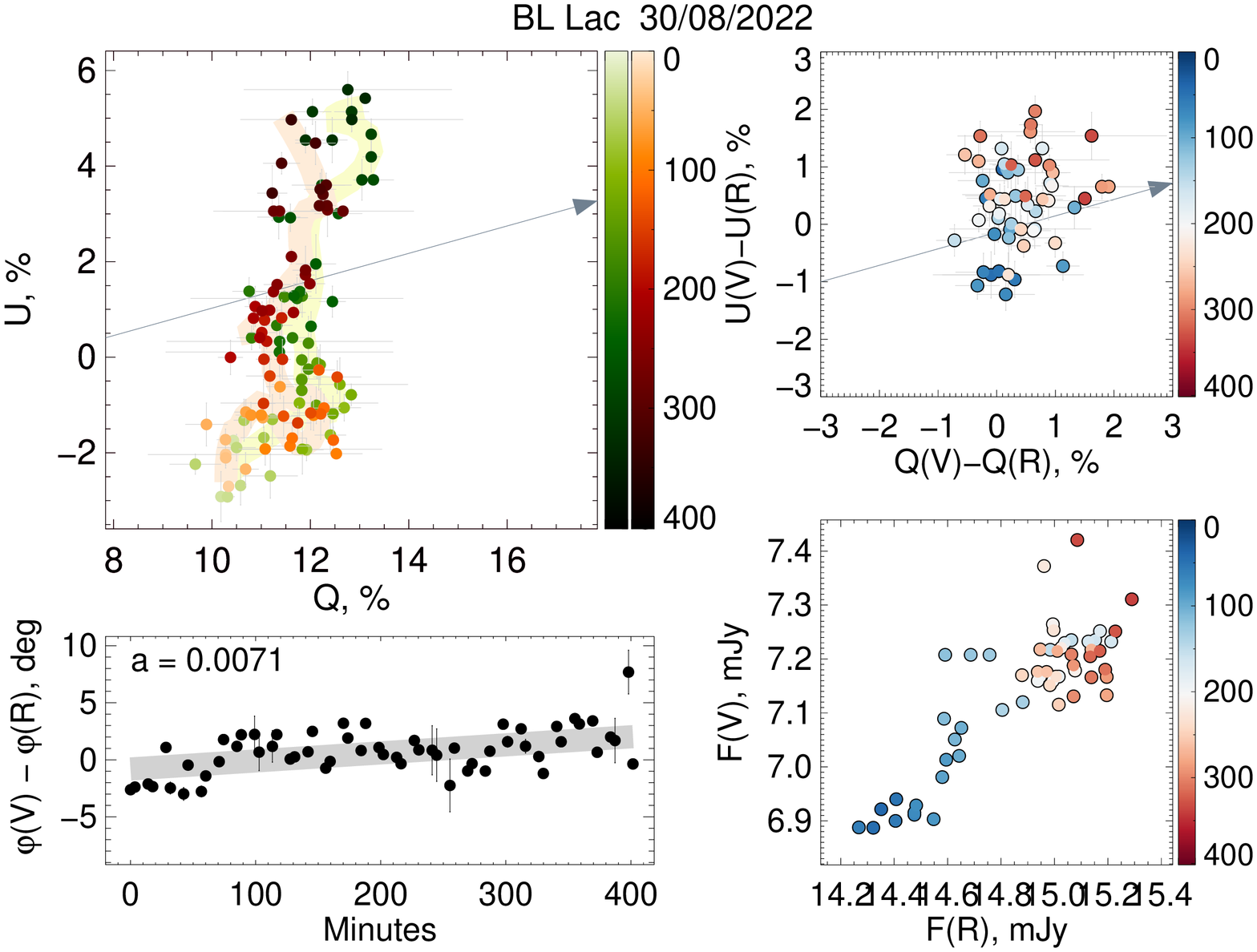}
    \caption{Similarly to Fig. \ref{A2} for the $V$ (green) and $R$ (orange-red) bands in the epoch 30/08/2022.}
        \label{A9}
\end{figure}

\begin{figure*}
     \centering
     \begin{subfigure}[b]{0.325\textwidth}
         \centering
         \includegraphics[width=\textwidth]{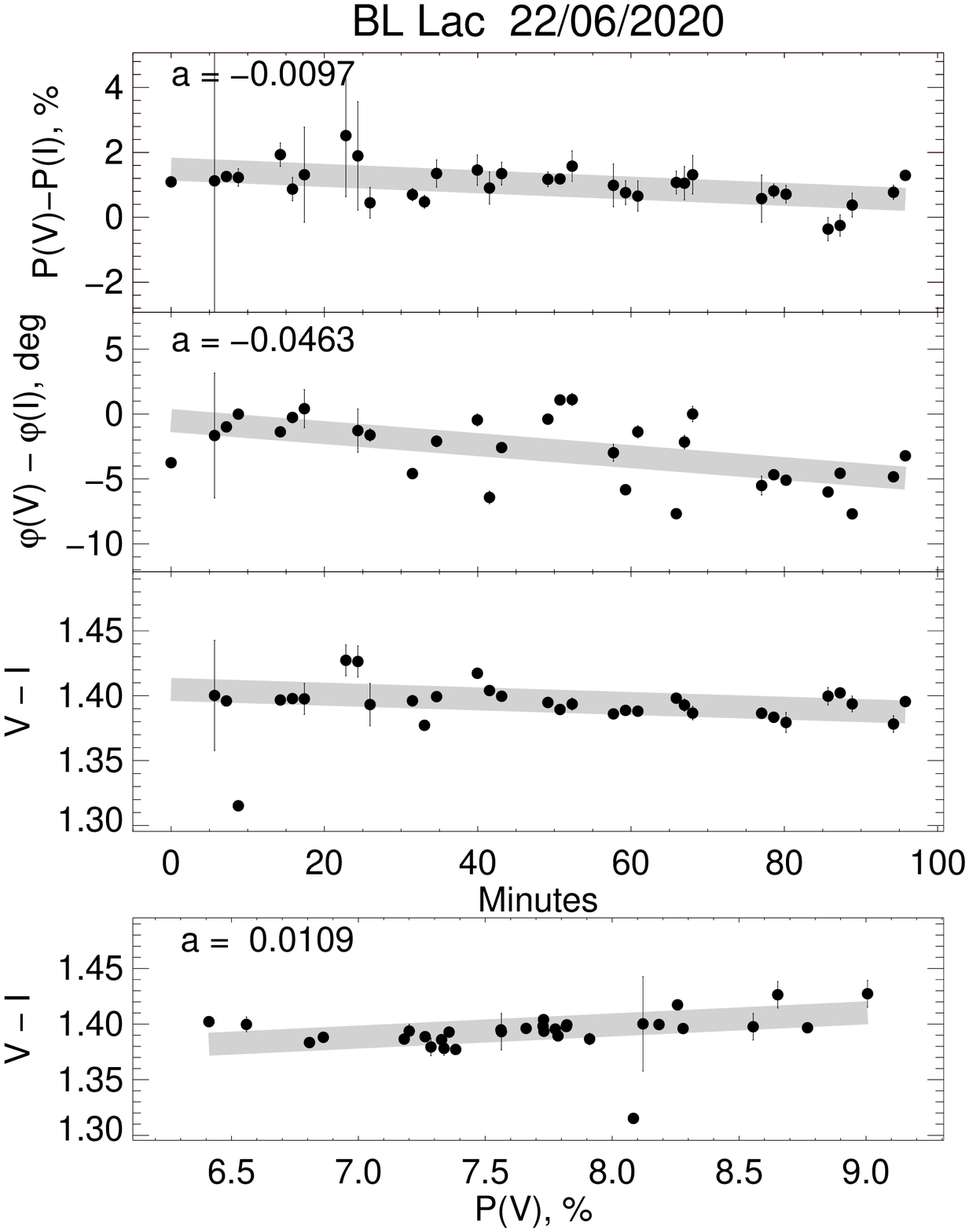}
     \end{subfigure}
     \hfill
     \begin{subfigure}[b]{0.325\textwidth}
         \centering
         \includegraphics[width=\textwidth]{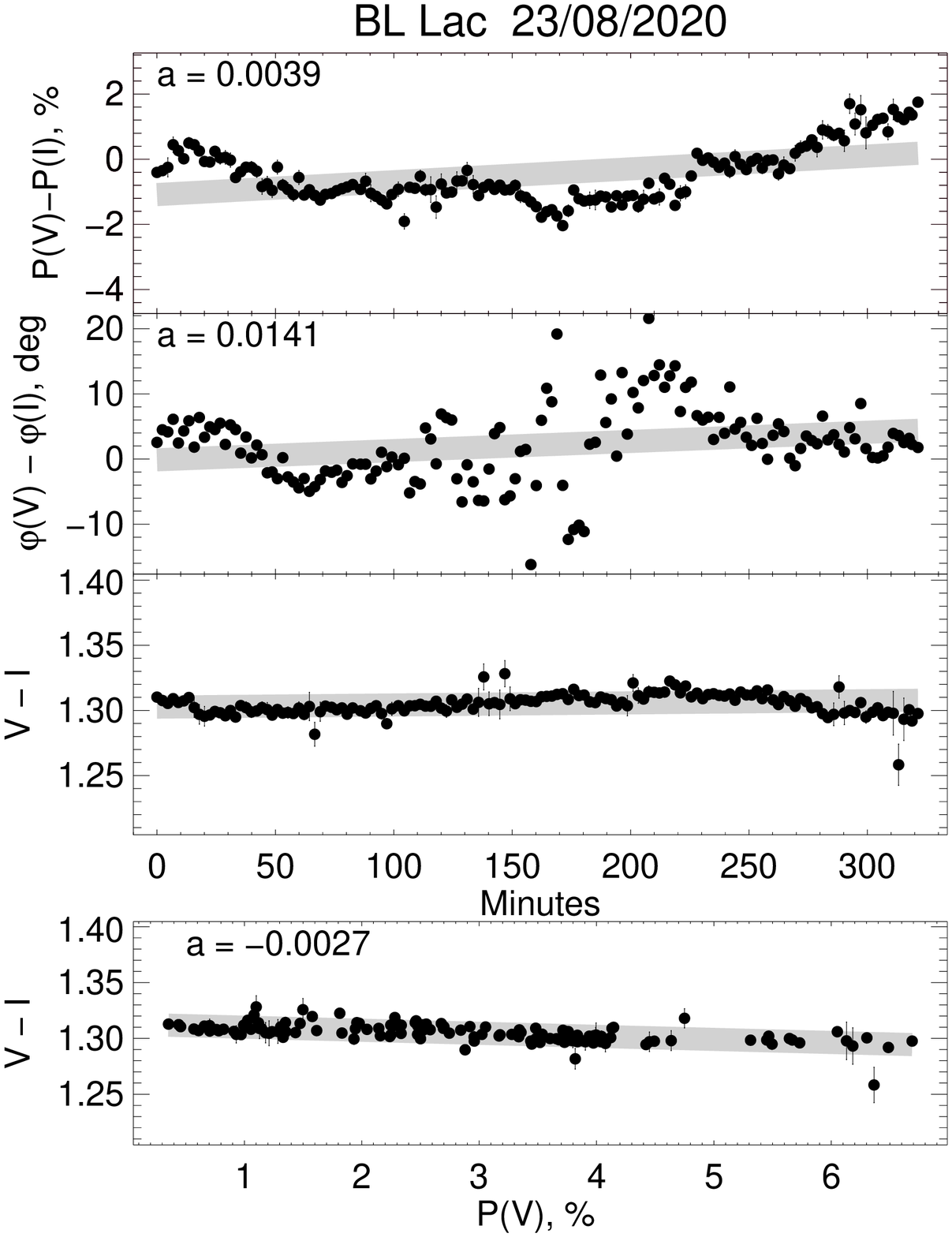}
     \end{subfigure}
     \hfill
     \begin{subfigure}[b]{0.325\textwidth}
         \centering
         \includegraphics[width=\textwidth]{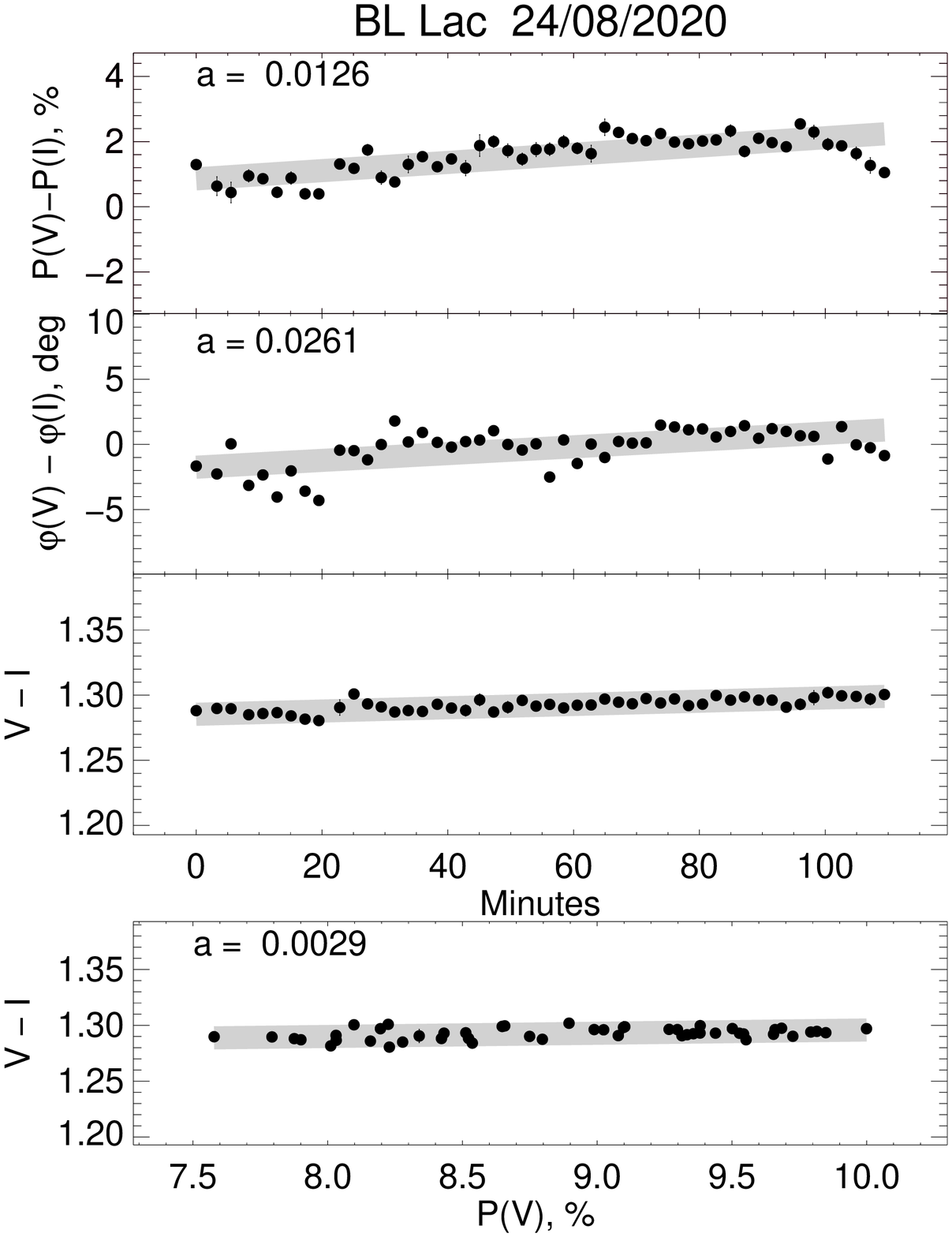}
     \end{subfigure}
    \hfill
    \begin{subfigure}[b]{0.325\textwidth}
         \centering
         \includegraphics[width=\textwidth]{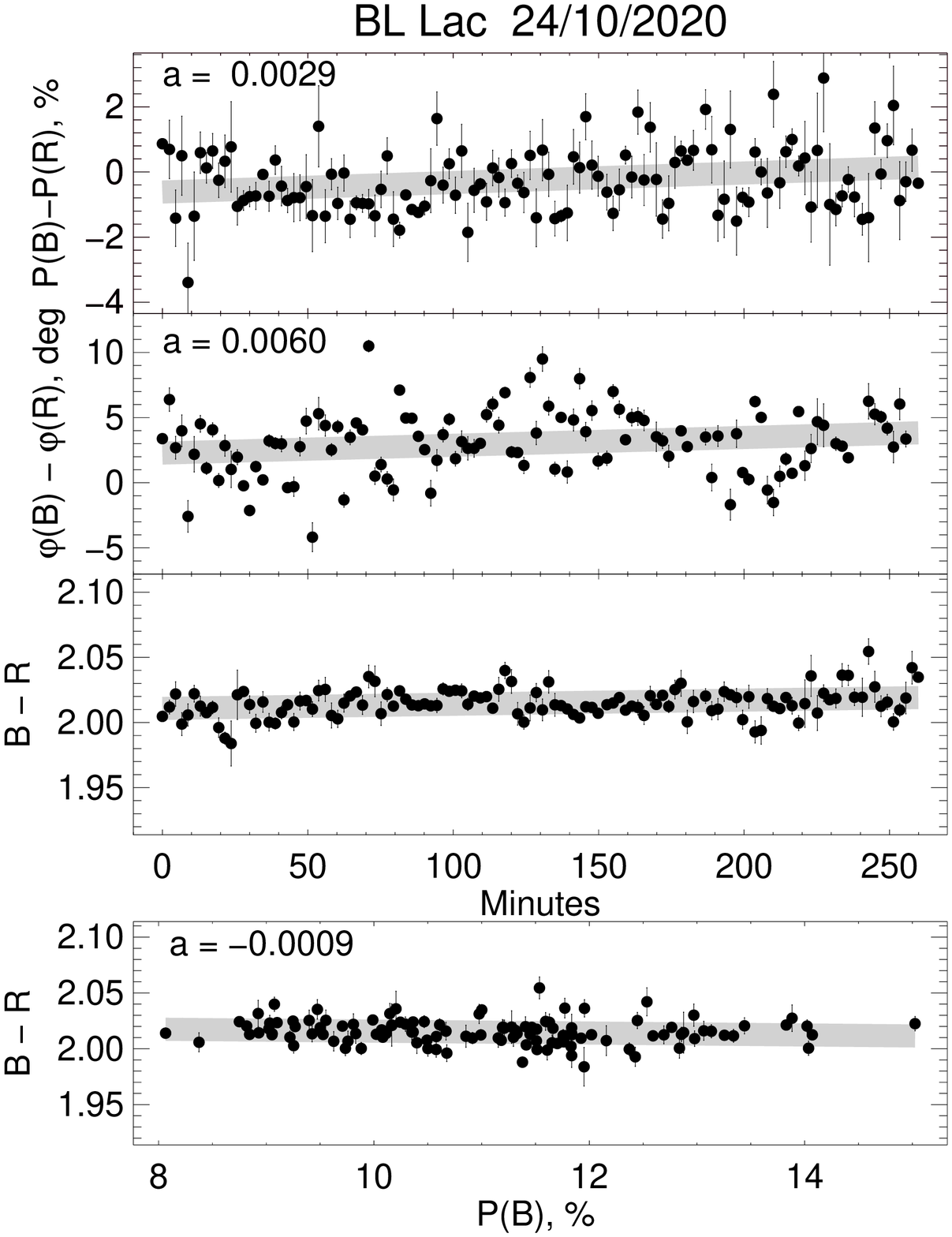}
     \end{subfigure}
     \hfill
     \begin{subfigure}[b]{0.325\textwidth}
         \centering
         \includegraphics[width=\textwidth]{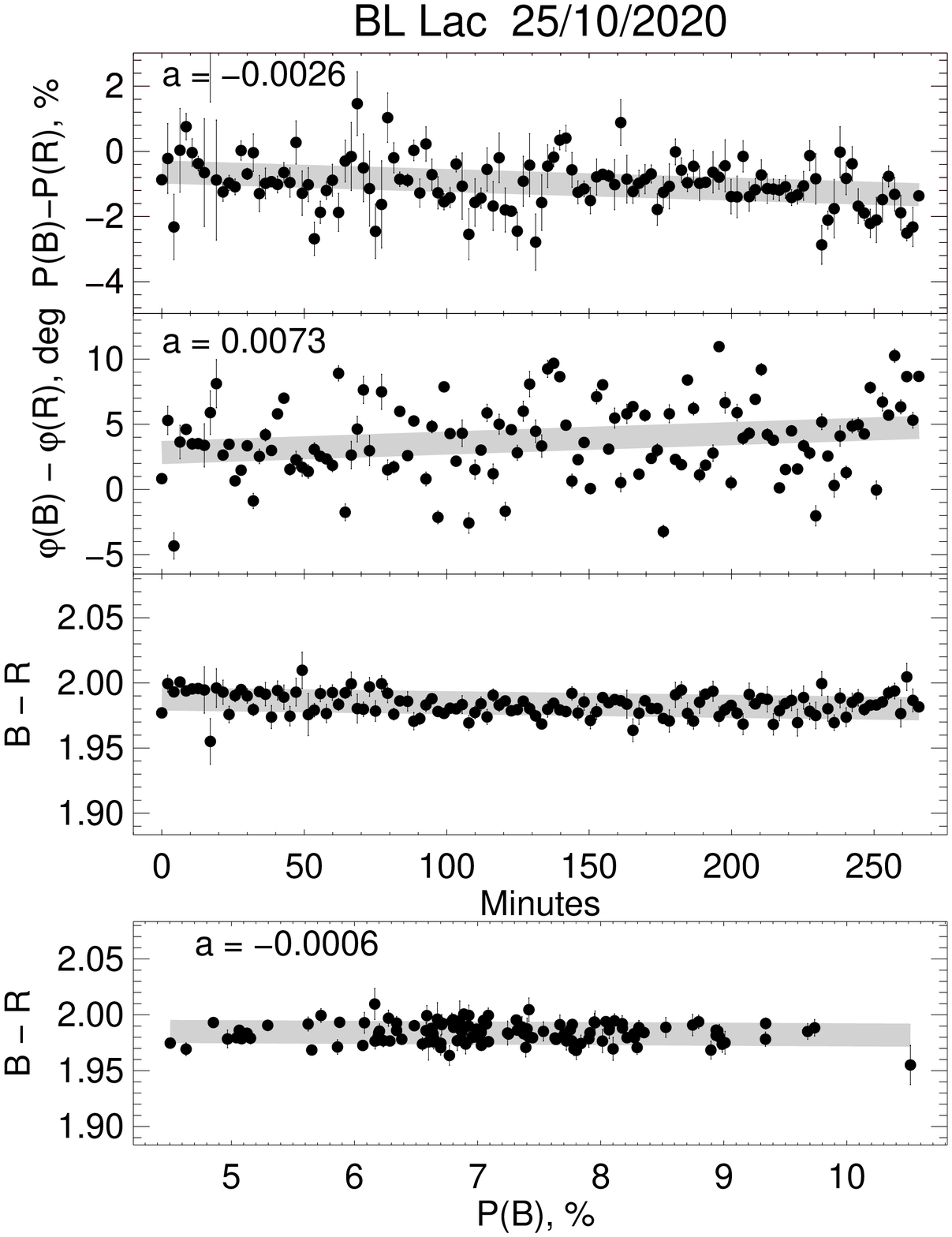}
     \end{subfigure}
     \hfill
     \begin{subfigure}[b]{0.325\textwidth}
         \centering
         \includegraphics[width=\textwidth]{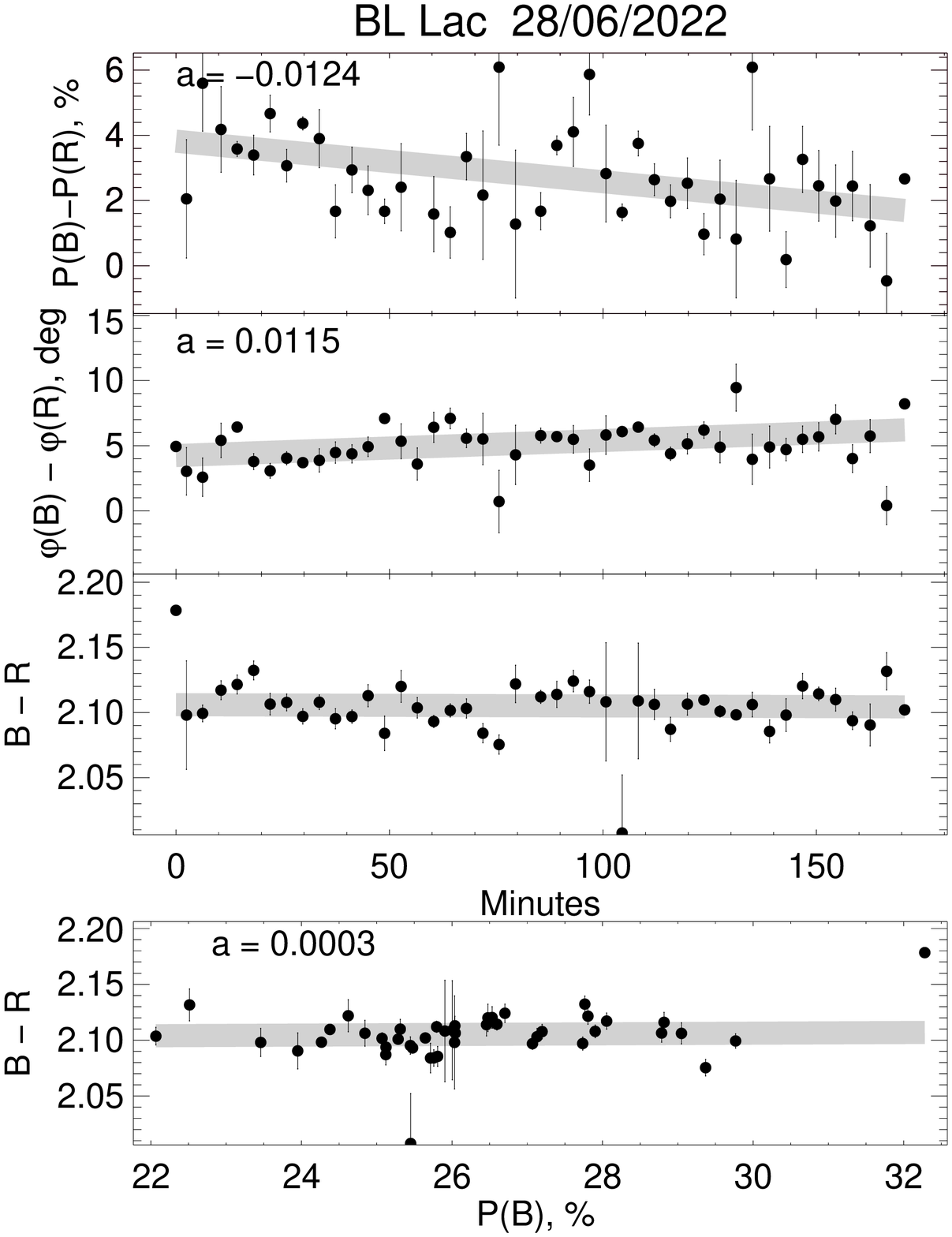}
     \end{subfigure}
          \begin{subfigure}[b]{0.325\textwidth}
         \centering
         \includegraphics[width=\textwidth]{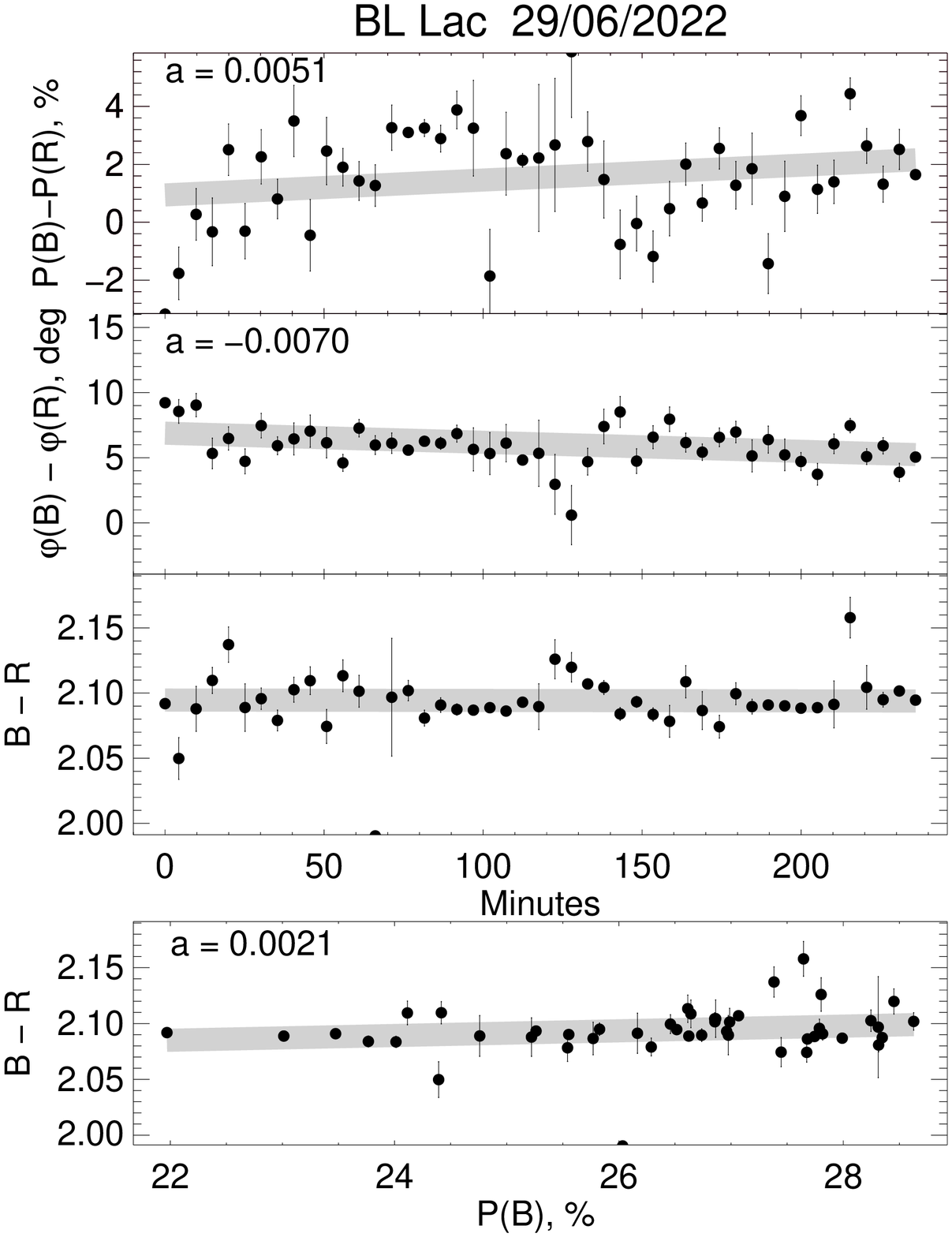}
     \end{subfigure}
     \hfill
     \begin{subfigure}[b]{0.325\textwidth}
         \centering
         \includegraphics[width=\textwidth]{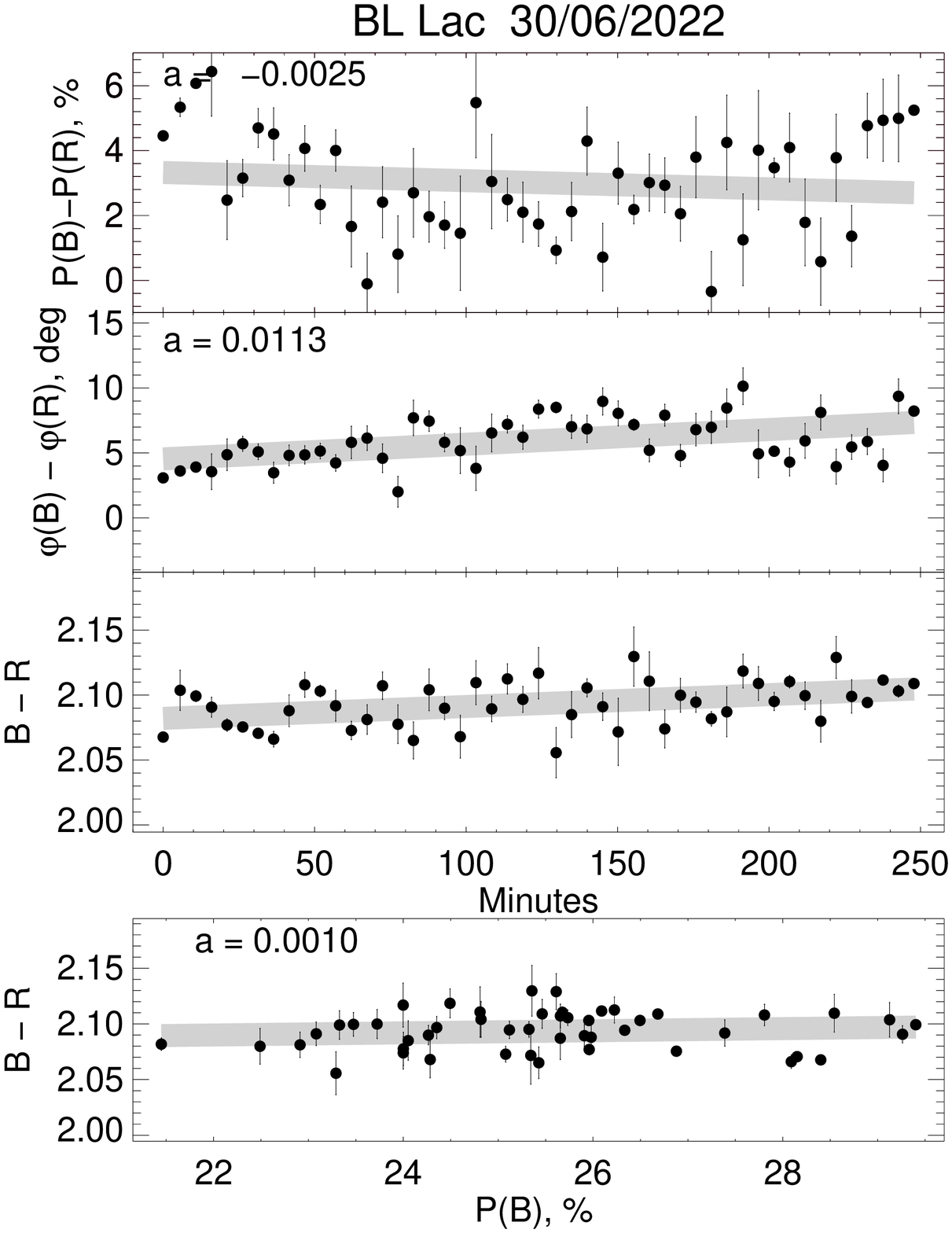}
     \end{subfigure}
     \hfill
     \begin{subfigure}[b]{0.325\textwidth}
         \centering
         \includegraphics[width=\textwidth]{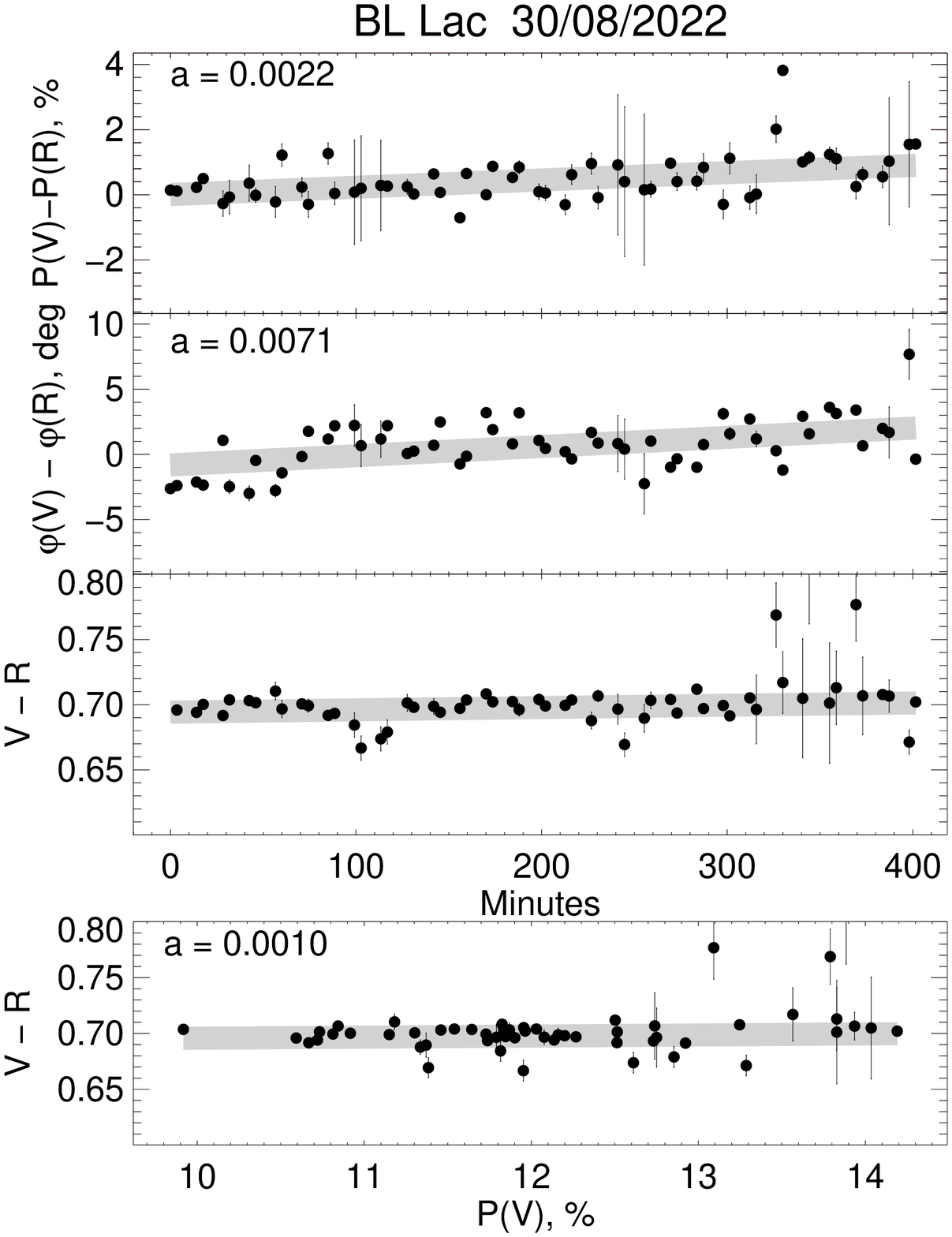}
     \end{subfigure}
        \caption{Light curves showing the difference in brightness and polarization between two colours. The panels correspond (from top to bottom): the difference in the polarization degree $P({\rm bluer}) - P({\rm redder})$, the difference in the polarization angle $\varphi({\rm bluer}) - \varphi({\rm redder})$, the difference in the magnitudes between the bands depending on the time in minutes, the difference in magnitudes between the bands depending on the polarization degree in the "bluer"{} band. Linear approximations of the data are given in grey lines. In the panels where indicated the parameter ${\rm a}$ describes the slope of the approximating line.}
        \label{pol_LC}
\end{figure*}

The section below presents additional figures demonstrating the observational data obtained in 2020-2022. In Fig. \ref{ap1} the light and polarization curves are presented, similar to Fig. \ref{LC}, for epochs when intraday polarimetric observations were carried out only in the $V$ filter. Fig. \ref{A2}-\ref{A9} are similar to Fig. \ref{fig:flareQU} for all other epochs not given in the main text of the paper. In more detail, the characteristics of the evolution of polarization chromatism during observational epochs as a function of time, as well as the evolution of the colour index from the degree of polarization, are given in Fig. \ref{pol_LC}.

\bsp	
\label{lastpage}
\end{document}